\documentclass{aa} 

\usepackage{time}
\usepackage[usenames]{color}
\usepackage{natbib}
\bibpunct{(}{)}{;}{a}{}{,} 

\def\setid#1,v #2 #3/#4/#5 #6 #7 #8.{(\textit{CVS version #2, checked in by #7 on #3-#4-#5, #6 UT})}
\usepackage{graphicx}
\usepackage[usenames,dvipsnames]{xcolor}
\usepackage{float}

\newcommand{\FeI}{\ion{Fe}{i}}
\newcommand{\CI}{\ion{C}{i}}
\newcommand{\kms}{\,km\,s$^{-1}$}
\newcommand{\ms}{\,m\,s$^{-1}$}
\newcommand{\fig}[1]{Fig.~\ref{#1}}
\newcommand{\figs}[2]{Figs.~\ref{#1} and \ref{#2}}

\definecolor{SanjColor}{rgb}{1,0,1}

\begin{document}

\authorrunning{Tiwari, van Noort, Lagg, Solanki}
\titlerunning{Sunspot penumbral filaments}

\title{Structure of sunspot penumbral filaments: a remarkable uniformity of properties}
\author{Sanjiv Kumar Tiwari$^1$, Michiel van Noort$^1$, Andreas Lagg$^1$, and Sami K. Solanki$^{1,2}$}
\date{Draft: \now\ \today\ \setid $Id: ms.tex,v 1.187 2013/05/27 09:32:26 sanjiv Exp $.}
\institute{$^1$Max-Planck-Institut f\"{u}r Sonnensystemforschung,
 Max-Planck-Str. 2, 37191 Katlenburg-Lindau.\\
$^2$School of Space Research,
Kyung Hee University, Yongin, Gyeonggi 446-701, Republic of Korea.\\
\email{[tiwari;vannoort;lagg;solanki]@mps.mpg.de}}
\offprints{S.~K. Tiwari \email{tiwari@mps.mpg.de}}

\abstract
{The sunspot penumbra comprises numerous thin, radially elongated filaments that are central for heat transport within the penumbra, but whose structure is still not clear.}
{We aim to investigate the fine-scale structure of these penumbral filaments.}
{We perform a depth-dependent inversion of spectropolarimetric data of a sunspot very close to solar disk center obtained by SOT/SP on board the Hinode spacecraft. We have used a recently developed spatially coupled 2D inversion scheme which allows us to analyze the fine structure of individual penumbral filaments up to the diffraction limit of the telescope.}
{Filaments of different sizes in all parts of the penumbra display very similar magnetic field strengths, inclinations and velocity patterns. The temperature structure is also similar, although the filaments in the inner penumbra have cooler tails than those in the outer penumbra.
The similarities allowed us to average all these filaments and to subsequently extract the physical properties common to all of them.
This average filament shows upflows associated with an upward pointing field at its inner, umbral end (head) and along its axis, downflows along the lateral edge and strong downflows in the outer end (tail) associated with a nearly vertical, strong and downward pointing field. The upflowing plasma is significantly, i.e., up to 800\,K, hotter than the downflowing plasma. The hot, tear-shaped head of the averaged filament can be associated with a penumbral grain. The central part of the filament shows nearly horizontal fields with strengths in the range of 1\,kG.
The field above the filament converges, whereas a diverging trend is seen in the deepest layers near the head of the filament. The fluctuations in the physical parameters along and across the filament increase rapidly with depth.}    
{We put forward a unified observational picture of a sunspot penumbral filament. It is consistent with such a filament being a magneto-convective cell, in line with recent MHD simulations. The uniformity of its properties over the penumbra sets constraints on penumbral models and simulations. The complex and inhomogeneous structure of the filament provides a natural explanation for a number of long-running controversies in the literature.}

\maketitle


\section{Introduction}

The complex filamentary nature of sunspot penumbrae has been the focus of numerous scientific investigations. Detailed reviews covering various open problems and discussing the proposed solutions have been given by \cite{sola03,thom04,thom08,scha09,trit09,borr09,schl09,bell10a,borr11,remp11}. 
One of the major open questions concerns the energy transport in penumbrae.
The strong magnetic field in penumbrae ($\sim$ 1 to 2\,kG) should prohibit convection according to various theoretical estimates \citep{cowl53,meye74,jahn94}. This would result in a much darker penumbra than observed. Obviously some form of convection must take place in the penumbra to account for its observed brightness. This convection can, however, take a number of different forms. 

Thus, the convection could occur in the presence of a strong magnetic field \citep{remp09a,remp09b,remp12}, indirectly supported by the observations of \cite{zakh08}, or in nearly field-free gaps separated by strong fields as proposed by \cite{spru06,scha06}.

The convection can also differ in the direction of the horizontal flows and the placement of the up- and downflows. Thus the horizontal flows could be mainly radial, with upflows located at the inner parts (heads, i.e., the portion close to the umbra) 
of penumbral filaments and ending as strong downflows at their outer parts (tails).
An example of such a flow is the Evershed flow \citep{ever09}. 
The flows may instead be mainly azimuthal, with upflows along the filament's major axis and downflows along its sides, or they may be some combination of the two. 
Some recent observations have found evidence for radial convection \cite[see e.g.,][and references therein]{rimm06,ichi07,fran09}. Indirect support for azimuthal convection was put forward by \cite{ichi07,zakh08,spru10,bhar10}, although an alternative explanation of some of these observations in terms of waves has also been provided \citep{bhar12}. Recent observations from the Swedish Solar Telescope (SST)/CRisp Imaging Spectro-Polarimeter (CRISP) showed downflows in the inner penumbra and a qualitative association of upflows with hot and downflows with cool gas in the deep photosphere \citep{josh11,scha11,scha12}. However some other researchers did not detect such flow patterns \citep{fran09,bell10,pusc10a}.

The relationship between the Evershed flow and the brightness of the penumbra is also not well understood. \cite{hirz01}, using 2-D spectrograms of high spatial resolution obtained from the Vacuum Tower Telescope (VTT: \citealt{schr85}), could not find a clear relationship between the Evershed flow velocity and the continuum intensity in the sunspot penumbra. However, \cite{titl93} and \cite{stan97} found that the Evershed flow is concentrated in dark filaments. A more detailed discussion on the above can be found in \cite{sola03}. 

Finally, there have been a variety of partially conflicting results on the complex magnetic structure of the sunspot penumbra and its relationship with the Evershed effect and the brightness pattern. Using observations from the VTT of a disc centered medium-size sunspot, \cite{schm92} found the field to be more inclined in the dark penumbral structures, whereas no variation of field strength between the dark and bright structures was detected. A similar conclusion was drawn by \cite{titl93} who used data obtained from the Swedish Vacuum Solar Telescope (SVST: \citealt{scha03}). In addition, they noticed the Evershed flow to be localized in the regions where the magnetic field was more horizontal. Soon after the above investigations, \cite{lites93}, using Advanced Stokes Polarimeter (ASP: \citealt{skum97}) data, discovered that the more vertical fields are stronger than their surroundings and named them 'spines'. However, they did not find a tight correlation between the intensity and the field inclination in the sunspot penumbra.
 
Only in partial agreement with \cite{lites93}, \cite{wieh00} reported that the dark penumbral features are associated with stronger and more horizontal magnetic fields. \cite{lang05}, however, found the opposite: the stronger penumbral field is more vertical and associated with brighter gas. More recently, \cite{borr11} noticed that the intra-spines (more horizontal and weaker fields) are brighter filaments in the inner penumbra and darker filaments in the outer penumbra. The magnetic structure in which spines (more vertical and stronger fields) and intra-spines are interlaced with each other has given rise to names such as ``uncombed penumbra''  \citep{sola93} or ``interlocking penumbra'' \citep{thom04}. \cite{sola93} also proposed that the more vertical magnetic field wraps around the more horizontal one. This appears to be the case according to Hinode data \citep{borr08,pusc10b,ruiz12}. 

Below we introduce the two most widely discussed proposed models describing penumbral magnetic structure. The first one is the embedded flux-tube model \citep{sola93}, or its dynamical extension, the rising flux-tube model \citep{schl98a,schl98b}. A siphon flow driven by gas pressure gradients due to different magnetic fields at the two footpoints of a flux tube is one proposed mechanism to drive the Evershed flow in this picture \cite[see e.g.,][]{meye68,thom93,mont93,dege93,mont97}. Another proposal is that the Evershed flow is driven by thermally induced pressure gradients along the flux tube \citep{schl98a,schl98b}. The presence of a significant pressure gradient along the Evershed flow channels has been deduced from the requirement of pressure balance with the surrounding magnetic fields by \cite{borr05}. The flux-tube models require localized upflows/downflows at the heads/tails of flux tubes. These models have successfully reproduced most of the spectropolarimetric observations, but have problems explaining the energy transport required to explain penumbral brightness along the full length of extended penumbral filaments \citep{schl03}. However, some recent investigations claim to have been able to clarify at least part of the penumbral brightness in terms of the Evershed flow based on the uncombed model of the penumbra \citep[see e.g.][]{ruiz08,pusc10a}.

The alternative model is the field-free gap model \citep{scha06,spru06}. According to this model, overturning convection takes place inside field-free gaps within the penumbra and transports an ample amount of heat to account for the penumbral brightness. This idea has similarities to the proposal by \cite{parker79a,parker79b}, see \cite{chou86}, to explain the brightness of umbral dots by means of intrusions of hot field-free gas into the umbra. A radial outflow corresponding to the Evershed flow may exist along the field-free gaps. However, the Evershed flow has been seen to be equally strong in Stokes $V$ as in Stokes $I$, as pointed out by \cite{sola94}, which implies that the flow takes place in a magnetized medium. Also, the structure of umbral dots obtained from simulations \citep{schu06} and observations \citep{riet08,riet13} indicates the presence of a magnetic field there. Given that neither of these proposals completely satisfies all constraints, we expect the true structure of penumbrae to combine aspects of both the models. 

Recent MHD simulations have been successful in reproducing many aspects of the fine scale structure of sunspots \citep{schu06,scha08,remp09a,remp09b,remp11a,remp12}. Magneto-convection is found to be responsible for penumbral fine structure and for driving the Evershed flow \citep{scha08,remp12}. Evershed flow channels are found to be strongly magnetized in deeper layers. \cite{remp12} finds azimuthal convection all along the penumbral filaments and finds that it is responsible for transporting the heat needed to account for the penumbral brightness. The magnetic field structure and its association with horizontal flows close to the solar surface in simulations shows similarities with the flux tube models, while the presence of azimuthal convection throughout the penumbra is close to the proposal of \cite{spru06}, although it is strongly magnetized.

It is widely accepted that the penumbra is made of copious thin filaments, well visible in high resolution intensity images \citep[e.g.,][]{dani61a,dani61b,beck69,moor81,wieh89,dege91,titl93,rimm95,west97,wieh00,lang05,ichi07a,borr08,borr11}. These filaments represent the main fine structure in sunspot penumbrae along with the dark spines \citep{lites93} and penumbral grains \citep{mull73,rimm06}. They also play a fundamental role in both models of penumbral fine structure described above. Therefore, it is important to determine their physical properties reliably. In the present paper, we investigate fundamental properties of the penumbral filaments of a sunspot observed almost at solar disk center. For this purpose, we isolated a significant number of penumbral filaments from all over the penumbra and divided them into three groups: filaments belonging to the inner, middle and outer parts of the penumbra, and compare the three in detail. 

The paper is structured as follows. In Section \ref{sec:obs} we describe our data and inversion technique as well as explain the selection criterion of the penumbral filaments. In Section \ref{sec:results}, results are presented. The results are discussed in Section \ref{sec:discussion} and appropriate conclusions are drawn in the last section.  

\section{Observational data and analysis techniques\label{sec:obs}}

We set out to study the structure of sunspot penumbrae observationally, in particular, we want to recover, if possible, any generic properties of the filaments that make up the penumbra. For this, we need to have high quality observations of a regular, quiet (i.e., non-flaring and only slowly evolving) sunspot, with as representative properties as possible, preferably at solar disc center to avoid any projection effects. The Stokes profiles from this sunspot then need to be inverted to obtain its atmospheric structure, using an inversion method that can recover the relevant atmospheric quantities as a function of height. Finally, due to the large variability in the appearance of the filaments caused by interactions with the environment as well as due to their intrinsic variability, we must find a way to identify and average the filaments, in order to study their generic properties.
\begin{figure*}[htp]
  \centering
  \includegraphics[width=0.82\linewidth]{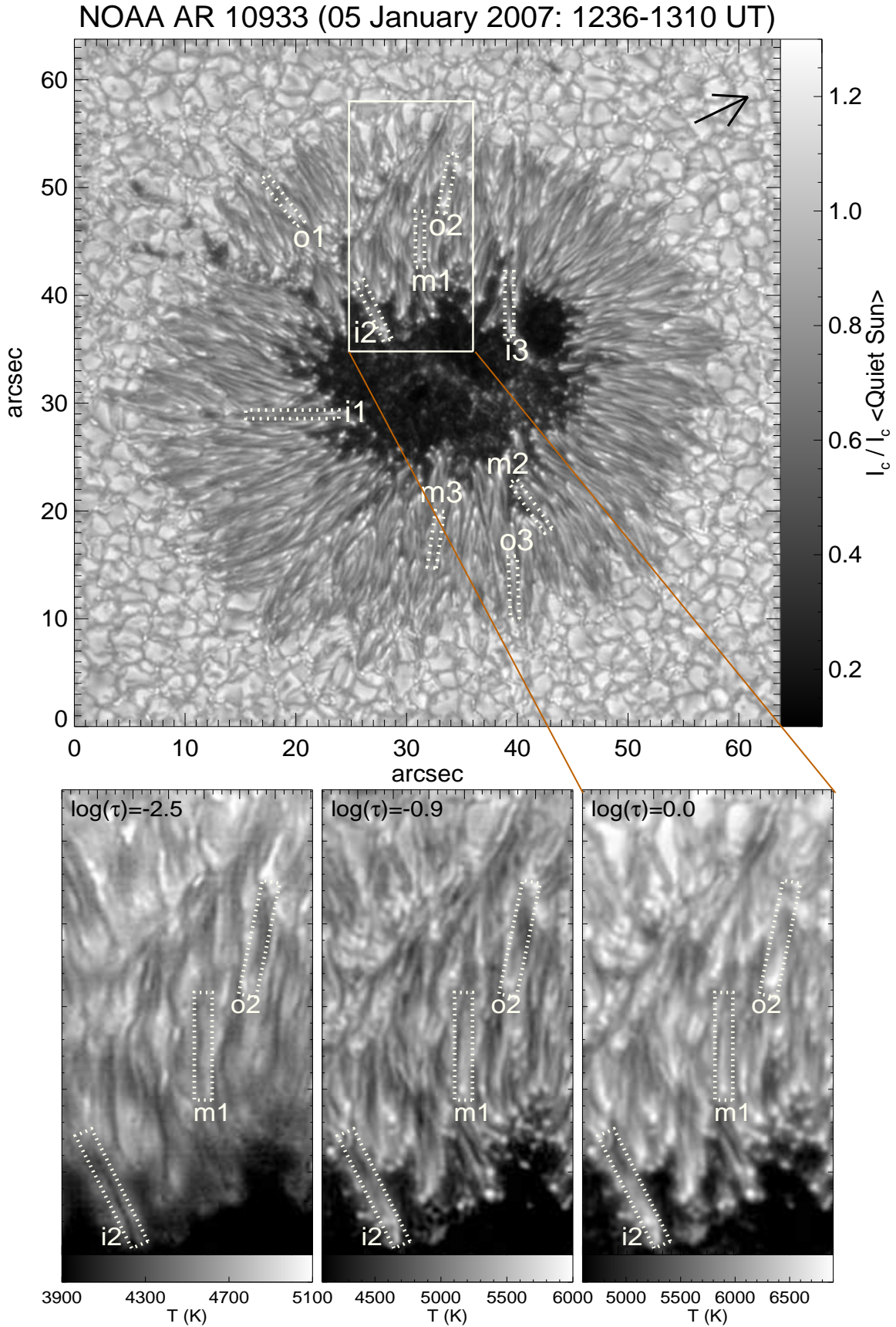}
  \caption{Upper panel: Continuum intensity map of NOAA AR 10933. The small dotted-line boxes represent locations of 9 (out of 60) penumbral filaments selected for study.  Filaments are not always ideally straight and therefore do not always fit completely in such boxes (which are for illustrative purposes only).  Three examples each of filaments located in the inner, middle and outer penumbra respectively are named as i1, i2, i3, m1, m2, m3 and o1, o2, o3. The arrow in the upper right corner points in the direction of the solar disc center. In the lower panels, we show temperature at $log(\tau) = -2.5$, $-0.9$ and $0$ for the region enclosed by the large solid-line box in the upper panel. Dark lanes within penumbral filaments are seen in the higher layers.}
 \label{fig:overview}
\end{figure*}

\subsection{Observations}
For our study, we selected the leading sunspot of NOAA AR 10933 that was observed in the \FeI\ 6301.5 and 6302.5\,\AA\ lines on the 5th of January 2007 from 12:36 to 13:10~UT using the spectropolarimeter of the Solar Optical Telescope (SOT/SP) \citep{tsun08,suem08,ichi08,shim08} on board the Hinode spacecraft \citep{kosu07}. The SOT/SP was operating in normal scan mode, resulting in a full-Stokes dataset with a spatial sampling of $\sim0.16$\arcsec{}/pixel, and a noise level of 10$^{-3}$ $I_c$ \citep{ichi08}. 

The sunspot umbra displays a positive magnetic polarity. It was observed almost exactly at solar disc center (S5$^\circ$, E2$^\circ$),
and was rather regular, with the exception of a small light bridge and a lane of enhanced brightness in the umbra. The data were calibrated by applying the standard data reduction routines available in the solar software package \citep{lites13}. Because of the nearly ideal characteristics of this particular sunspot, different aspects of it have been investigated by several researchers \citep[see e.g.,][]{fran09,venk10,kats10,fran11,borr11}. 

\subsection{Inversions}
We infer the atmospheric properties of the spot by inverting the observational data. For the inversion of the data, we made use of the spatially coupled version \citep{van12} of the SPINOR inversion code \citep{frut00}. The code builds on the STOPRO routines \citep{sola87} to solve the radiative transfer equations for polarized light \citep{unno56,rach67} under the assumption of local thermodynamic equilibrium (LTE). 

The inversion code calculates the best fitting full-Stokes spectra of an atmospheric model described by a selected number of atmospheric parameters, specified at a number of optical depth positions and interpolated using a bicubic spline approximation. The spatially coupled version is able to invert the observed data taking into account the spatial degradation introduced by the telescope diffraction pattern, allowing for a consistent and accurate fit of the observed spectra, while keeping the atmospheric model as simple as possible. The atmospheric parameters returned by the code are the best fits to the Stokes profiles in the absence of the blurring effect of telescope difraction. They show small-scale structures sharper than in the original data to the extent allowed by noise etc.

In this paper we make use of spatially coupled inversions described in detail by \cite{downflowpaper}, where the inversion is carried out on a denser grid than that of the original data. The resulting inverted result has a significantly improved spatial resolution in most fit parameters and appears to produce a more robust inversion result than at the original pixel size. Here we follow \cite{downflowpaper} and use a pixel size of 0.08''/pixel, a factor 2 smaller than in the original data, allowing structures down to the diffraction limit of the telescope to be adequately resolved.

The inverted parameters are temperature $T$, magnetic field strength $B$, field inclination $\gamma$, field azimuth $\phi$, line-of-sight velocity $v_{los}$ and a micro-turbulent velocity $v_{mic}$. All the parameters were allowed to vary at all three height nodes, which were placed at $\log(\tau)= 0, -0.9$ and $-2.5$ respectively.

To encourage the inversion to find the best match to the data, multiple cycles of inversion, followed by an ad-hoc smoothing of the fitted atmosphere were executed (explained in detail by Tiwari et al., in preparation). The smoothing is intended to ``disturb'' the solution, which is provided as an initial guess for next iteration of the inversion. This procedure was repeated until no further improvement of the merit function is obtained. The atmosphere returned at this point is considered to be the one providing the best fit to the data.

If we change the node positions, and/or add different levels of noise in the original and convolved profiles, the inverted quantities are very similar in extent and manifestations. Such tests were performed by \cite{downflowpaper} to establish the robustness of the inversion. In addition, a good agreement between the inversion results and MHD simulations for the velocities and fields was found \citep{van12,downflowpaper}, further supporting the reliabilty of our inversion process.

However, it is difficult to precisely establish the error in the fitted atmospheric parameters. Because of spatial coupling, any change in a parameter at a pixel can result in a variation of the same parameter in a neighbouring pixel. Thus, it is necessary to compute the solution for a perturbation of each parameter at every pixel simultaneously. This will need a large amount of computational resources and will be presented in detail in another publication. In this paper, therefore, we refrain from presenting error estimates of the inversion.

After the inversion, the 180$^\circ$ azimuthal ambiguity was resolved using the minimum energy method \citep{metc94,leka09}.

An offset of 100 \ms{} was subtracted from the line-of-sight velocity. This offset was determined by assuming that the umbra, excluding the umbral dots, is at rest on average. An independent velocity calibration following \cite{rodr11} on the average profile of the \FeI\ line at 6301.5012 \AA\ over the least active parts of the Sun contained in the Hinode SP scan returned a velocity offset of approximately 200 \ms{} of the calibrated data with respect to the gravitational redshift corrected FTS atlas \citep{brau87}. This is slightly more than the offset of 100 \ms{} that was removed and may indicate that no truly quiet Sun was contained in the scan and the offset may be attributed to the presence of magnetic elements in the selected regions, as outlined by \cite{bran90}.

\subsection{Filament selection, de-stretching and length-normalization\label{sec:select}}

\begin{figure*}[ht]
  \centering
  \includegraphics[width=0.7\linewidth]{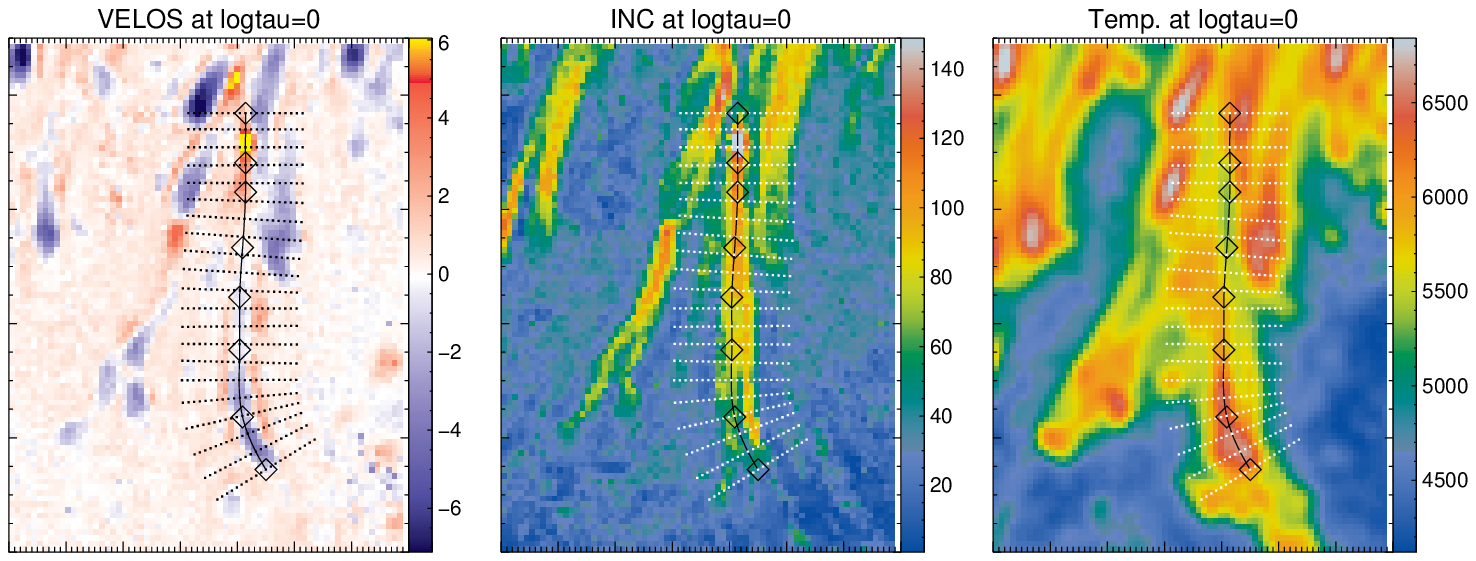}
  \includegraphics[width=0.7\linewidth]{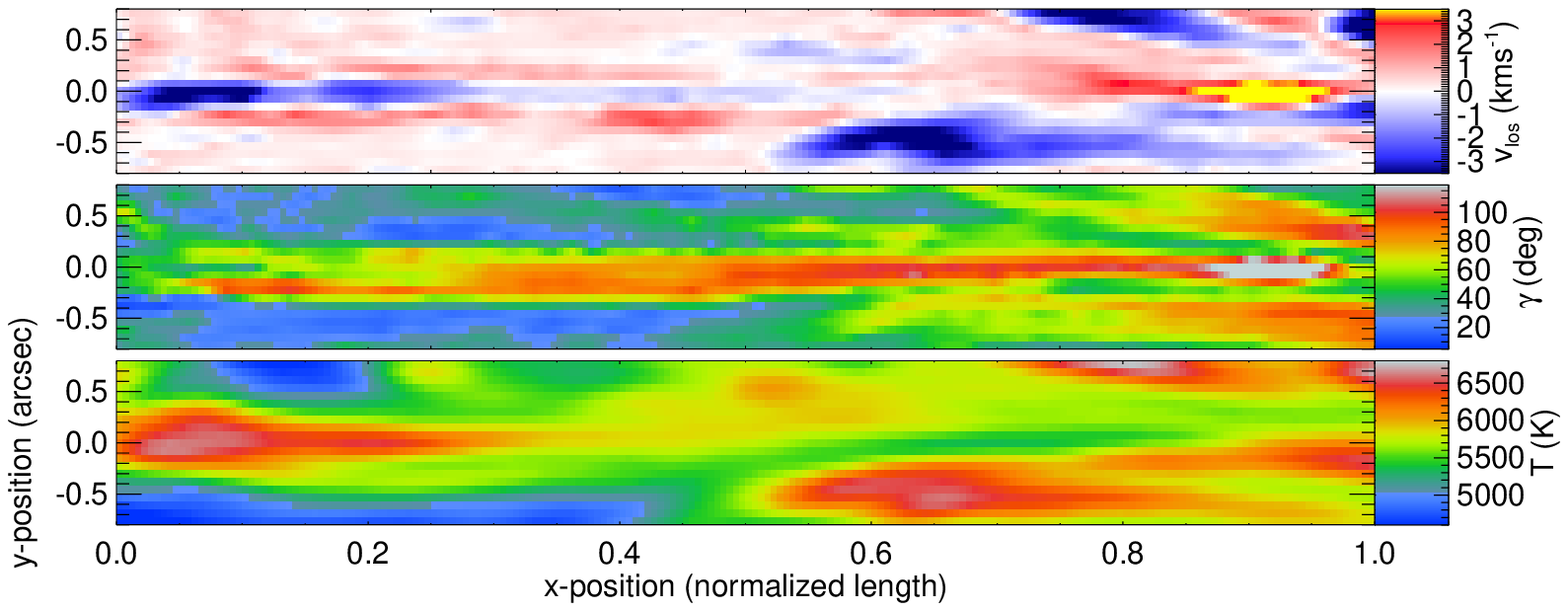}
  \caption{Selection of a filament. The three maps from left to right in the upper panel display line-of-sight velocity v$_{los}$ (in \kms{}), field inclination $\gamma$ (in degrees) and temperature $T$ (in K). The diamonds are selected points lying along the filament's axis between which a spline interpolation is performed.  The area selected for the study of this filament is covered by the transverse lines. The v$_{los}$, $\gamma$ and $T$ of the selected filament in its de-stretched and scaled form at log($\tau$)=0 are shown in the lower panel.}
  \label{fig:tracking}
\end{figure*} 

As we are interested in the properties of filaments, we must first find a method to identify and isolate the filaments.

Figure \ref{fig:overview} (upper panel) shows the continuum image, synthesized from the best fit to the data. Despite the large amount of detail visible in the image, the isolation of individual filaments from the intensity image alone is rather ambiguous, as the filaments are often not straight, show poor contrast to the surroundings (particularly in the outer penumbra), and sometimes merge and overlap with other filaments, making it sometimes hard to decide which feature belongs to which filament.

One problem is that a fraction of filaments is complex. Such complex filaments have either multiple heads feeding a single tail or a single head from which multiple filament bodies extrude. In order to avoid complications, we concentrate here on simple filaments, i.e. those with a single head connected to a single tail.    

We have found it to be most reliable to identify filaments using multiple parameters returned by the inversions. Besides the continuum intensity we employ the line-of-sight velocity and the inclination of the magnetic field, both at optical depth unity. Our criteria (see below) work for most filaments because many of them turn out to show a qualitatively similar structure. The heads of the filaments are clearly identifiable by a teardrop-shaped, upflow of some 5--10 \kms{}, with the pointed ``tail'' of the drop (not to be mistaken with the tail of the filament) pointing radially outward. The region immediately behind the drop usually contains only weak up- and downflows and is not particularly suitable for isolating the main body of the filaments, while the tails of filaments can again be well identified in velocity since they display substantial, concentrated downflows. The body of a filament is best traced by using the inclination of the magnetic field, which allows the heads and tails of the various filaments (seen, e.g., in velocity) to be in general uniquely assigned to each other.

In the upflow regions at the heads of the filaments the field points upwards i.e., in a direction similar to that of the umbral field. It is nearly horizontal in the body of the filament and points downwards in the filament tails, co-located with the strong downflows present there. This inclination ``lane'' is typically flanked by lanes of relatively vertical field, aligned with the umbral field, corresponding to the so called spines \citep{lites93}. In a few ambiguous cases, other inverted parameters were also used to keep track of the filaments.

This procedure was applied manually by placing points along the middle of the filament, which turned out to be more reliable than our attempts at an automated identification. A line connecting these points was then computed using a bicubic spline interpolation, as illustrated by the black line in \fig{fig:tracking} (upper panel), which follows a filament that protrudes into the umbra. The path defined in this way defines the (possibly curved) axis of the filament and was used in the de-stretching and length normalization of the filament. The tangent of the path was subtracted from the azimuth angle, effectively aligning the filament with the azimuth reference direction. In addition, the azimuth value on the central axis of the filament was taken as an offset and removed, so that the value of the azimuth along the major axis is always zero. This correction was necessary to remove the background caused by a global twist in the sunspot's magnetic field (which is $<$ 5$^\circ$, see e.g., \citealt{tiw09b}) and any remaining calibration errors in the data. 

To perform the de-stretching and length normalization, 200 points were placed equidistantly along the axis of each filament, after which 10 points on a line perpendicular to the tangent are placed at intervals of 1 pixel ($0.08$'') on each side of the path. A cubic interpolation of the inverted parameters in these points results in a de-stretched, de-rotated and length normalized filament, as shown in the lower panel of \fig{fig:tracking}. 

\section{Results}
\label{sec:results}
\subsection{Size of filaments} 
With the method described in the previous section, a total of 60 penumbral filaments were selected and de-stretched, 20 each from the inner, middle and outer parts of the sunspot penumbra. The inner penumbral filaments originate close to or even inside the umbra, middle filaments are those which start in the inner to middle parts of the penumbra and end well inside the outer penumbral boundary. The outer filaments originate in the middle to outer regions of the penumbra and end near or at the outer penumbral boundary. In \fig{fig:overview} a number of the inner, middle and outer filaments are highlighted by dotted-line boxes.

\begin{figure}[h]
  \centering
  \includegraphics[width=0.82\linewidth]{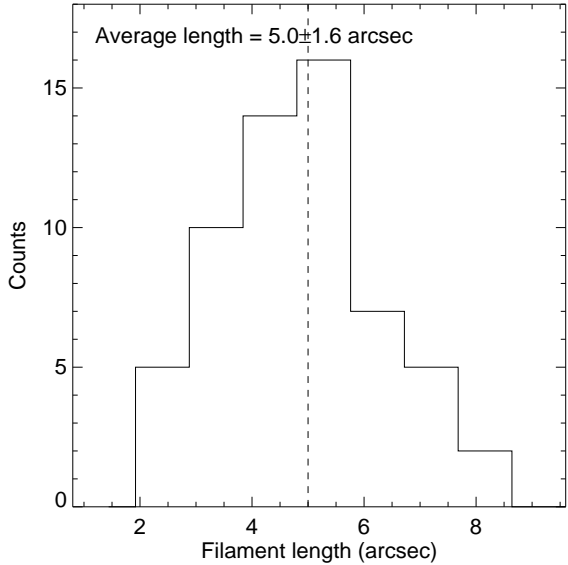}
  \caption{Histogram of the lengths of all the 60 selected penumbral filaments.}
  \label{fig:hist}
\end{figure} 

The selected filaments exhibit a wide range of lengths, therefore making re-scaling necessary for averaging, as described earlier in the Section \ref{sec:select}. The width of the filaments is however not re-scaled. To have an estimate of the length and width of the filaments, we make a histogram of the actual lengths of all the 60 filaments and display in \fig{fig:hist}. We find that the mean length of the filaments is about 5\arcsec{} with a standard deviation of $\pm$1.6\arcsec{} around the mean. The actual width after averaging all filaments comes out to be 0.8\arcsec{}. Thus, from the average of the measured lengths of the individual filaments, we arrive at a length-to-width ratio of approximately 6. There was no significant difference in the average lengths of the filaments selected in the inner, middle and outer penumbra.

On the one hand, while averaging, the variation in the length will cause some properties to be re-scaled along the filament. On the other hand the generic physical properties and structure, reflected from the average of such filaments, will be valid for filaments of all sizes.  

\begin{figure}
  \centering
  \includegraphics[scale=0.73]{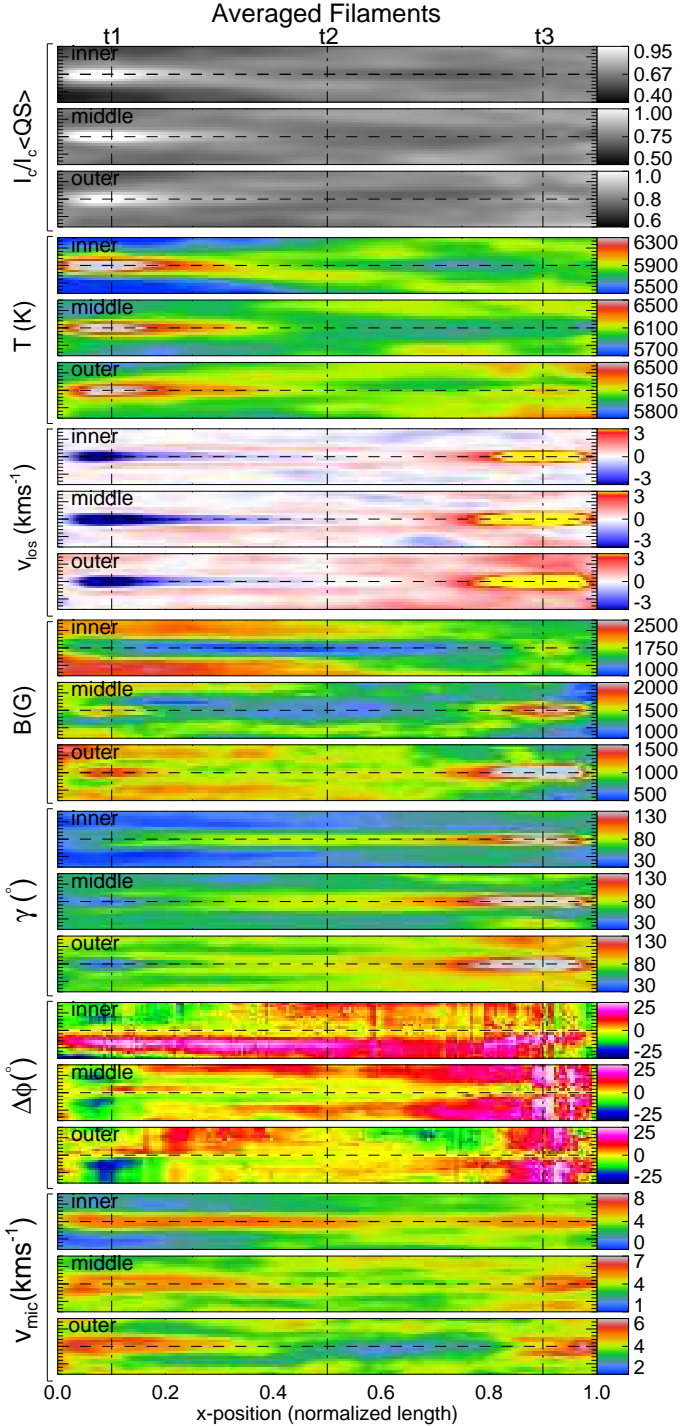}  
  \caption{From top to bottom: 3 plots each of the physical quantities fitted continuum intensity $I_c/I_c<QS>$, where $<QS>$ implies the average over all quiet Sun pixels, temperature $T$, line-of-sight velocity $v_{los}$, magnetic field strength $B$, inclination $\gamma$, azimuth $\Delta\phi$ and microturbulent velocity $v_{mic}$ of averaged filaments from inner, middle and outer parts of the sunspot penumbra. $X=0$ refers to the head of filament (the position closest to the umbra). The radial location in the penumbra of each panel is labelled on the upper left corner of each panel. All plots are at depth $log(\tau) = 0$. The dashed and dot-dashed lines mark a longitudinal and three transverse cuts (labelled as t1, t2, t3 above the top panel, belonging to head, center and tail of the filaments, respectively) along which various physical parameters are plotted in \figs{fig:long_cut}{fig:trans_cut}. Note that color bars for a single parameter do not always have the same range for all three maps of a given physical parameter. The color scale of the velocity has been saturated at $\pm 3$ \kms{} in order to highlight the smaller velocities along the bulk of the filament.}
  \label{fig:avgd_maps}
\end{figure}

\subsection{Qualitative picture of filaments}

A look at the inversion results suggests that filaments (at least a heavy majority of them) have many properties in common. These include: 
\begin{enumerate}
\item Fluctuations in all parameters, both along and across a filament, increase strongly with depth, with filaments being most clearly distinguishable in the lowest atmospheric layers (but see one exception in point \ref{p:darklane} below).
\item The brightness of a filament is largest at its head and decreases towards its tail, in agreement with the findings of \cite{bhar10}, although many have a second enhancement in the tail (in the location of the downflows).
\item \label{p:darklane}Most filaments do not display central lanes in the continuum intensity or $T$ at $log(\tau)=0$, but show increasingly marked dark lanes in $T$ at $log(\tau)=-0.9$ and $-2.5$, in agreement with the interpretation that the dark lanes, first reported by \cite{scha02}, arise where canopy-like magnetic fields meet and close.
\item The field strength also drops somewhat from the head to the body of a filament, again with a local strengthening associated with the downflows at the very tail.
\item The magnetic inclination of filaments looks like a strongly flattened $\bigcap$-loop, with a very elongated horizontal part. In particular, all filaments seem to show opposite polarity fields at their tails, not just those at the outer edge \citep[e.g.,][]{west97}.
\item Filaments also appear to have a qualitatively similar velocity structure, with upflows concentrated at the head but continuing in a thin band along their central axis, so that the upflowing region has the general appearance of a teardrop-headed pin. The downflows tend to lie along a strongly elongated V-shape that surrounds the upflows on three sides, although for most filaments the two arms of the V have different strengths of downflow. In a number of cases one arm completely disappears.
\end{enumerate}
 
These qualitatively similar properties of basically all simple filaments led us to devise the procedure for their identification and isolation (Section \ref{sec:select}) and to consider averaging them in order to reduce the still significant fluctuations from one filament to another. 

\subsection{Properties of filaments in the inner, middle and outer penumbra\label{sec:properties}}

The 20 selected filaments from each penumbral region (inner, middle and outer), destretched and length-normalized as outlined in section \ref{sec:select}, were then averaged to recover their generic properties. The averaging of the filaments reduces random errors, introduced while selecting the individual filaments, as well as intrinsic variations of the individual filaments themselves, revealing those properties they have in common.

\begin{figure}[ht]
  \centering
  \includegraphics[width=\linewidth]{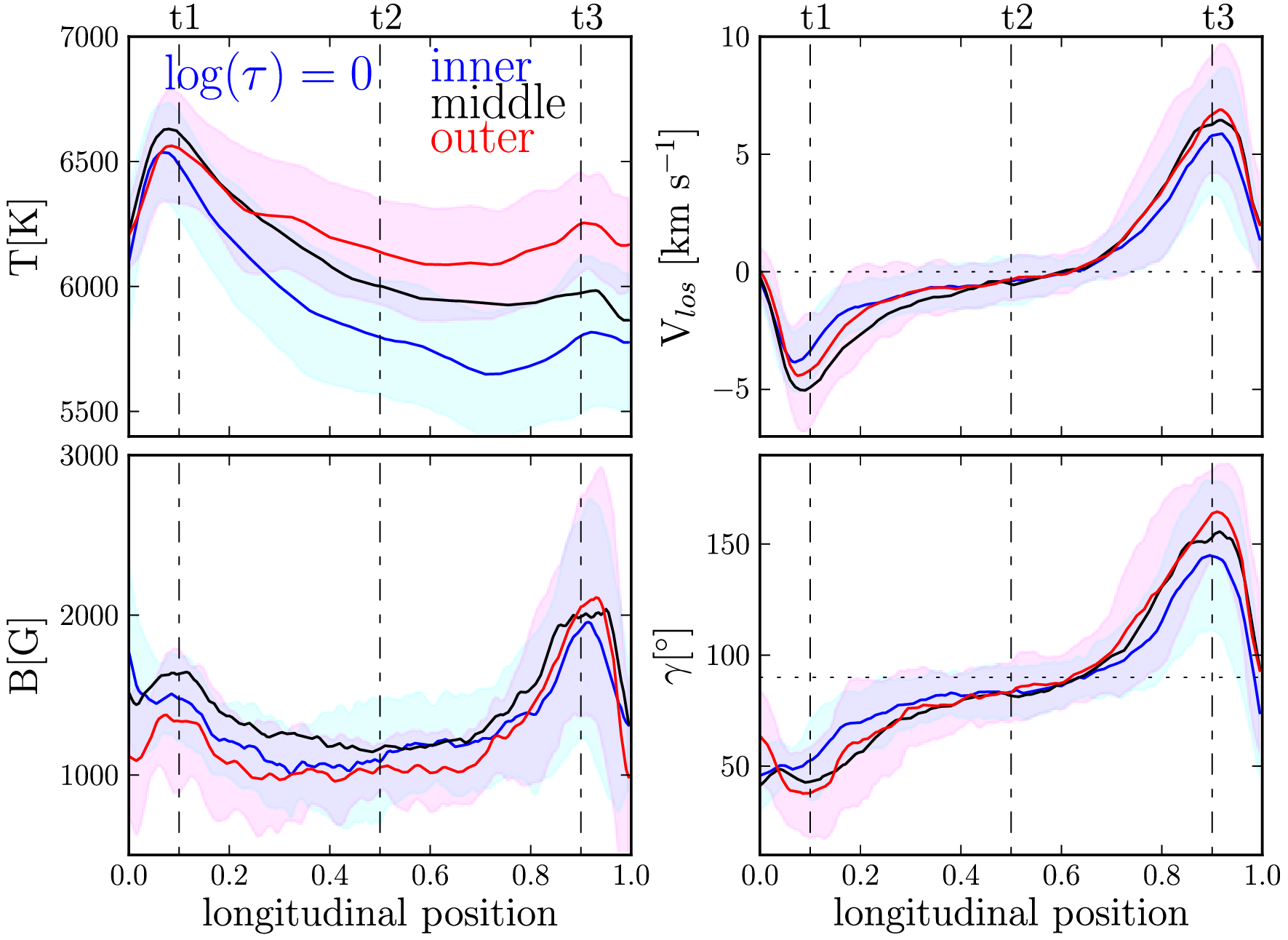}
  \caption{Variations of the physical parameters $T$, $v_{los}$, $B$, and $\gamma$ at $log(\tau)=0$ along the longitudinal cut denoted by horizontal dashed lines in \fig{fig:avgd_maps}. Parameters of averaged filaments in the inner, middle and outer parts of the penumbra are plotted together in each panel. $v_{los}$=0 and $\gamma=90^\circ$ are indicated by horizontal dotted lines. Locations of the cuts t1-t3 are marked as vertical dot-dashed lines. The standard deviations of the parameters among the 20 filaments entering the averaged inner and outer curve are indicated by the semi-transparent shaded areas, with each having the same color as the line it is associated with. The violet shading indicates overlap between properties of the filaments in the inner and the outer penumbra. The standard deviation of the averaged middle filament is not shown to keep the picture clearer.}
  \label{fig:long_cut}
\end{figure}

\begin{figure*}[ht]
  \centering
  \includegraphics[width=0.98\textwidth]{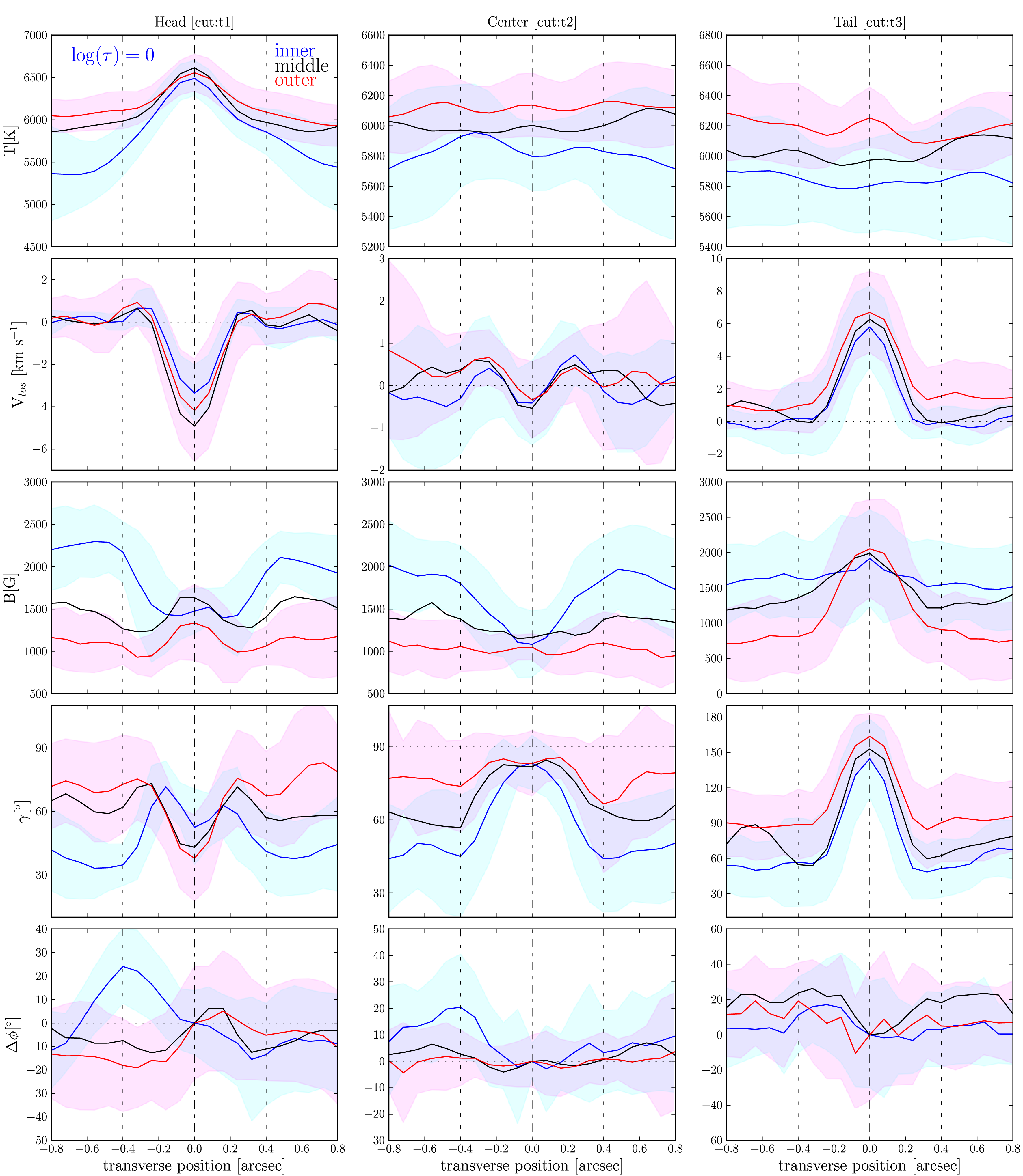}
  \caption{Same as \fig{fig:long_cut} but for cuts across the filaments at the three given transversal cut positions 't1', 't2' and 't3' belonging to the head, center and tail of filaments respectively.  Two vertical dashed lines represent the boundary of the filaments. $v_{los}$=0, $\gamma=90^\circ$ and $\Delta\phi=0^\circ$ are indicated by horizontal dotted lines.}
  \label{fig:trans_cut}
\end{figure*}

Figure \ref{fig:avgd_maps} depicts a selection of properties of each of the three averaged, standardized filament, one representative for each penumbral region. Note that the filaments are slightly narrower than the frame, so that a bit of the (averaged) surroundings can also be seen for each. The symmetry of the averaged quantities with respect to the central axis is striking for all three averages and for all selected quantities. In addition, although only 20 filaments were averaged, the remaining fluctuations in the averages, with the possible exception of the averaged azimuth angle, are quite small compared to the overall patterns, suggesting that fluctuations around the average behaviour are uniform and relatively unimportant, at least for the filaments selected here (c.f. the standard deviation shown in \figs{fig:long_cut}{fig:trans_cut}). These figures display cuts of some of the quantities along the filament's axes and perpendicular to them, respectively.

Although there are differences between the three averaged filaments (as discussed below), the similarity is striking. This strongly suggests that although the appearance of the penumbra varies from the inner to the outer penumbra, the mechanisms at work are remarkably uniform.

In addition to the obvious presence of strong upflows, the heads of the 
filaments at optical depth unity are characterized by enhanced temperature (and/or continuum intensity) and a relatively vertical and enhanced magnetic field. A feature that stands out clearly in all the three averages is the weak downflow channel surrounding the upflow on three sides.

An interesting feature is that the temperature in the heads of all three averaged filaments is rather similar, while the temperature in the surroundings of the head increases significantly with radial distance of the filament. This can also be clearly seen in \fig{fig:trans_cut} along cut ``t1'' (near the head). The continuum intensity (not shown in the plots of \figs{fig:long_cut}{fig:trans_cut}), as expected, shows a very similar behaviour to temperature for all the three averaged filaments. 

Similarly, the magnetic field strength surrounding the head of the inner penumbral averaged filament is significantly stronger than inside the filament head, but decreases rapidly in strength at larger distances from the center of the spot (see also \citealt{borr05}), to comparable or even lower field strength than in the head itself. As clearly revealed by \figs{fig:long_cut}{fig:trans_cut}, the field strength in the filament itself is remarkably independent of the location of the filament. A closer inspection of the inclination angles shows a similar picture, with near vertical inclination surrounding the inner filaments gradually changing to more horizontal. The magnetic inclination of the filaments themselves is, again, very similar for all the three averaged filaments.  

Clearly, the head itself does not change much, but the environment it is found in shows a significant and systematic variation as a function of the distance to the umbra. The change in the properties of the surroundings is partly due to the variation of the spines with radial distances and appears to have a significant impact on the downstream temperature structure of the filaments. Whereas the inner filaments show a rapid decrease of the temperature with the distance to the head, as can be seen in \fig{fig:long_cut}, the further away from the umbra the filament starts, the slower the temperature decreases, so that the bodies and tails of the outermost filaments remain some 500\,K warmer than of their inner counterparts.

However, none of the other atmospheric parameters seems to be significantly affected, as can be seen in \fig{fig:long_cut}. Line-of-sight velocity, magnetic field strength and inclination show only modest differences between filaments at different radial distances. Any remaining observed differences may well be caused entirely by density changes associated with the changes in the temperature structure. Even the width of the filament, does not show any dependence on their radial distance from the umbra, with all averaged filaments showing peripheral downflow channels at $\pm$5 pixels ($\pm$0.4'') from the filament axis.

A consistent feature in the maps of the line-of-sight velocity at optical depth unity (\fig{fig:avgd_maps}) is the appearance of a narrow downflow channel on each side of the filament, that retains a fairly uniform strength along the entire filament and converges slowly to form a strong downflow at the end of the filament. The upflow that defines the head of the filament extends to more than half the length of the filament, before it gives way to a downflow.
 
Compared to their surroundings, the magnetic field in the filaments is weaker, specially in the inner penumbra but still $\geq$1\,kG everywhere in the filaments, with quite significant enhancements at the head and tail. In the enhanced areas, the field is more vertical, the head displaying the same polarity as the umbra, while the field in the tail displays the opposite polarity. In particular the downflows in the outer penumbra are associated with nearly vertical fields. The enhancements in the magnetic field strength are clearly spatially correlated with the peak values of the up- and downflows.

The field azimuth angle, as defined in section~\ref{sec:select}, shows an approximately antisymmetric behavior with respect to the filament axis, but is not consistent between the different filaments, and is rather noisy (see also \fig{fig:trans_cut}). The noisy character of the plots suggests this quantity is not particularly well constrained in the inversion, at least at this height. It is, however, more stably and consistently deduced at higher layers (the depth dependence is described below).

\begin{figure}[ht]
  \centering
  \includegraphics[width=\columnwidth]{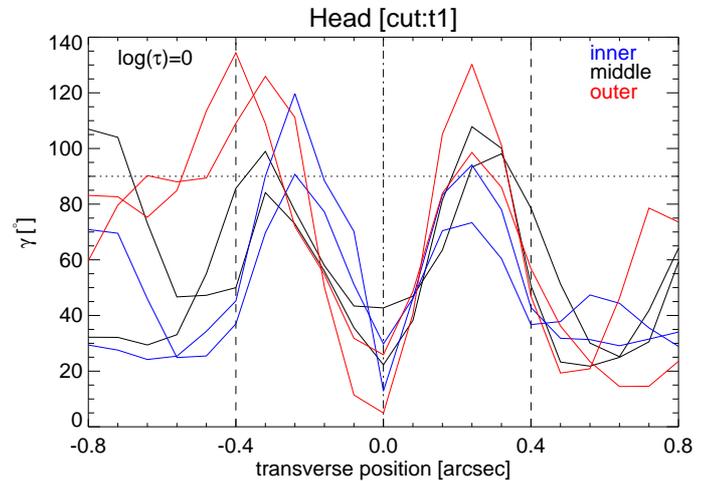}
  \caption{Field inclination along a transverse cut in location t1 of six individual filaments, two of each belonging to the inner, middle and outer penumbra, showing opposite polarity field ($\gamma > 90^\circ$) near their edges on at least one side. The two vertical dashed lines indicate the boundary of the filaments used when forming the averages. The horizontal dotted line denotes $\gamma=90^\circ$.}
  \label{fig:incl_opp_pol}
\end{figure}

In the transversal direction, near the head of the filament (along cut ``t1'', \fig{fig:trans_cut}), the field is somewhat weaker and more horizontal at the location of the downflow channels. The increase in inclination persists up to the center of the filament, until it finally disappears near the tail. The decreased magnetic field strength is clearly visible only near the head.

From the fact that the shaded region extends to $\gamma > 90^\circ$ in the middle-left panel of \fig{fig:trans_cut}, we expect that at least some of the selected filaments should show opposite polarity ($\gamma > 90^\circ$) in location t1. A few of these filaments showing opposite polarity on at least one side are plotted in \fig{fig:incl_opp_pol}. In total, some 30\% of all filaments show opposite polarity along their edges on at least one side. There appears to be a trend that the opposite polarity fields at the sides of the filaments near their heads get stronger and more vertical for the filaments located further from the umbra. We also noticed that the opposite polarity fields occur at different distances from the axes of the various filaments. This means that when averaging over them, these opposite polarities get smoothed out rather effectively.

Comparing the line-of-sight velocity to all other quantities yields a clear correlation of the flow speed with the inclination, field strength, continuum intensity and/or temperature. All quantities show a local extremum at the locations of ``t1'' and ``t3'': whereas the head is dominated by a peak in temperature, enhanced ``umbral'' polarity magnetic field and a significant upflow, the tail shows a much lower temperature enhancement, strongly enhanced opposite polarity field and a strong downflow. The downflow is on average stronger than the upflow.

As seen from \fig{fig:avgd_maps}, the microturbulence is clearly higher along the filament's major axis than in the surroundings, mainly in the inner penumbra. In the middle and outer averaged filaments, $v_{mic}$ shows enhancements only near the extended heads and tails. This suggests that the filaments or at least parts of them host strong, unresolved velocities. 

\subsection{Convective nature of filaments}

\begin{figure}[h]
  \centering
  \includegraphics[scale=0.8]{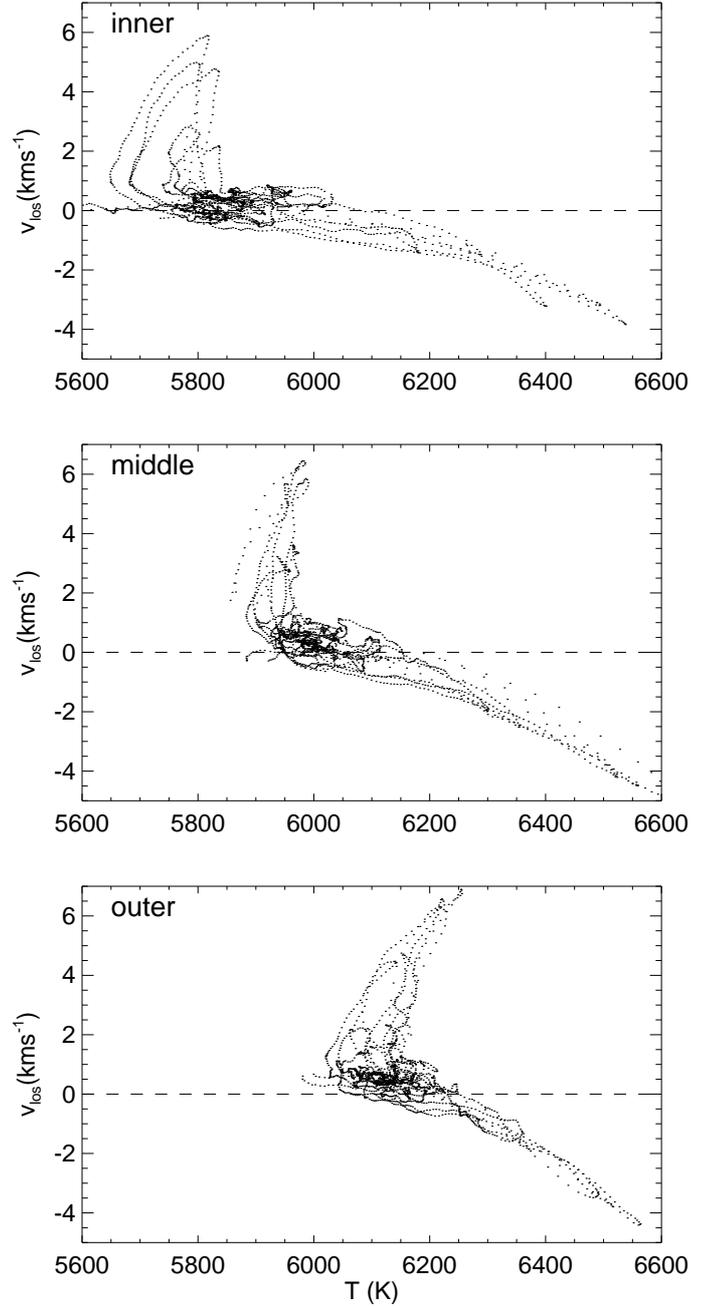}  
  \caption{From top to bottom: scatter plot between the temperature and line-of-sight velocity of the three averaged filaments belonging to the inner, middle and outer penumbra, respectively. Note that the surrounding pixels (more than 0.4\arcsec{} away from the filament's central axis) are not taken into account. The alignments of the points along curves are the result of the spline interpolation along individual filaments when producing the averaged filament. This process can distribute similar values onto several pixels. The horizontal dashed line in all the three panels marks the zero velocity level.}
  \label{fig:scat}
\end{figure}

To explore the convective nature of penumbral filaments, we display in \fig{fig:scat} scatter plots of the temperature vs. the line-of-sight velocity in the three averaged filaments belonging to the inner, middle and outer parts of the sunspot penumbra. Pixels more than 0.4\arcsec{} away from the filament's central axis (i.e., those showing the surroundings) are not taken into account. As can clearly be seen in the figure, the upflows (i.e., negative velocities) are systematically hotter than the downflows (i.e., positive velocities) for all three averaged filaments, with this difference reaching values of up to 800\,K in the inner penumbral filaments.   
The increasing temperature of the downflows with radial distance in the spot can clearly be seen in the three scatter plots.  

The hook-like shape outlined by the downflowing data points in \fig{fig:scat} suggests that the faster downflows are on average somewhat warmer than the slower downflows. This behaviour is particularly clearly visible for the averaged outer penumbral filaments.

As we shall see in Section \ref{sec:massflux} the flows all start in the bright and hot head 
of the filament, where the field is more vertical, and continue up to more than half of the filament along its central axis. Material is then carried outward along the filament's axis (Evershed flow) and across it in the azimuthal direction. Along the way, the gas cools and sinks again at the tail as well as along the sides of the filament.

\subsection{Standard filament}

The strong indication from the previous section that all filaments have essentially the same structure, motivates us to average all filaments to produce a single,  ``standard'' filament, describing the general properties of filaments found anywhere in the penumbra. This makes sense only for the penumbral filaments themselves, as the rather substantial variation of the environment would not justify averaging. Moreover, the continuum intensity and temperature of such a standard filament is unique only near the head, as the downstream value depends on the penumbral position, as found in Section \ref{sec:properties}.

\begin{figure}[ht]
  \centering
  \includegraphics[scale=0.67]{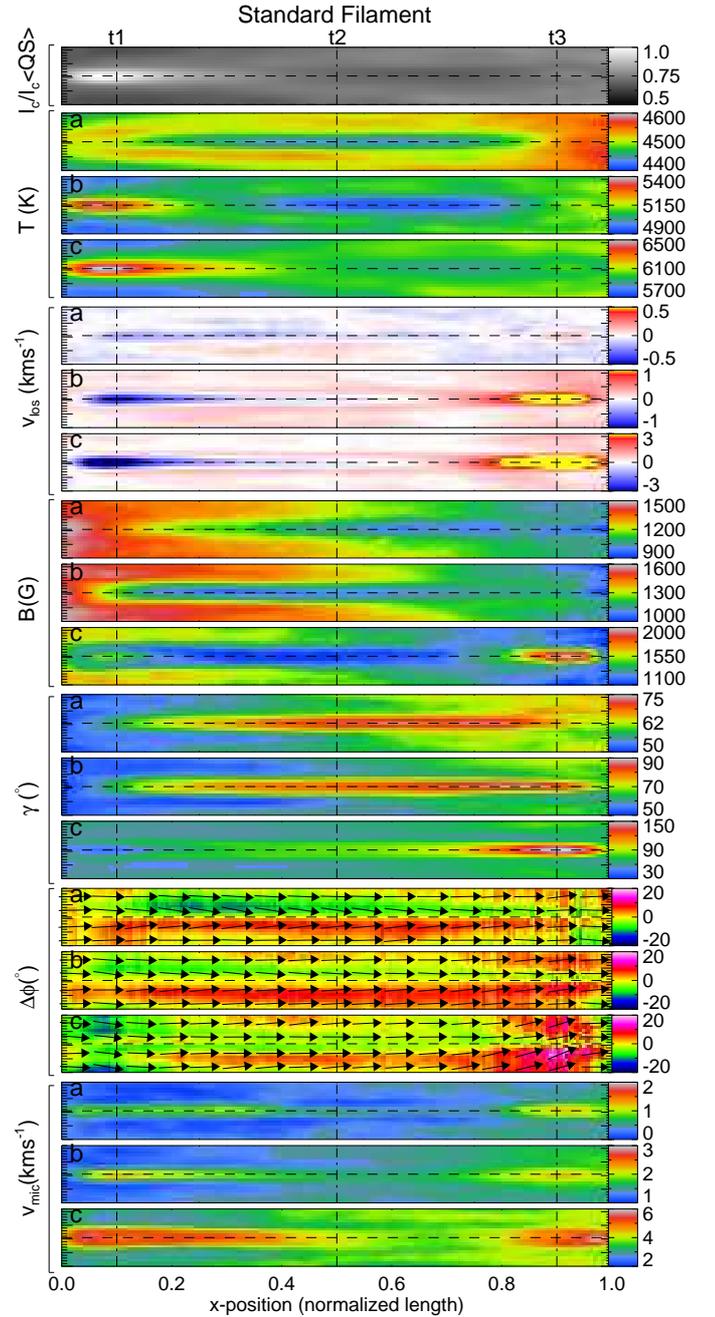}  
  \caption{Horizontal slices of the physical parameters of the ``standard'' filament, resulting from averaging all 60 selected filaments, at three optical depths, including the fitted continuum intensity map as the topmost panel.  The labels ``a'', ``b'' and ``c'' correspond to $\log(\tau)=-2.5, -0.9$ and $0$, respectively. The vectors overplotted on the offset-removed azimuth maps are meant to guide the eye only.  The one horizontal dashed and three vertical dot-dashed lines are the positions of the longitudinal and transversal cuts shown in \figs{fig:slice_long}{fig:slice_trans}, respectively.}
  \label{fig:stand_fil}
\end{figure}

\fig{fig:stand_fil} displays the result of averaging the inversion outcomes of all 60 selected filaments from all over the penumbra, for $\log(\tau)= -2.5, -0.9$ and $0$. Given the similarities between the filaments in the different parts of the penumbra, it is not surprising that this globally averaged or ``standard'' filament shows an overall structure very similar to that obtained for each penumbral region separately.

\begin{figure}[ht]
  \centering
  \includegraphics[scale=0.52]{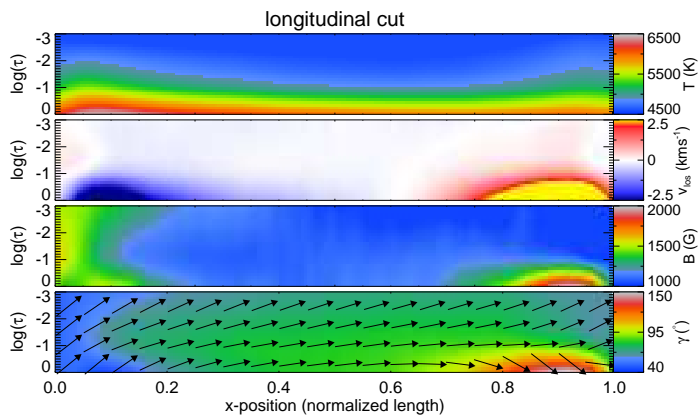}  
  \caption{Longitudinal-vertical slice through the standard penumbral filament along the central axis. The line-of-sight velocity is saturated at $\pm2.5$ \kms{} in order to show weaker flows as well. The arrows in the $\gamma$ map indicate the direction of the magnetic field, mainly to guide the eye. However one must keep in mind that, among other things, the optical depth corrugation effect is not taken into account. Thus the geometrical height needs not be the same at the head and tail of the filament.}
  \label{fig:slice_long}
\end{figure}

\begin{figure}[ht]
  \centering
  \includegraphics[scale=0.58]{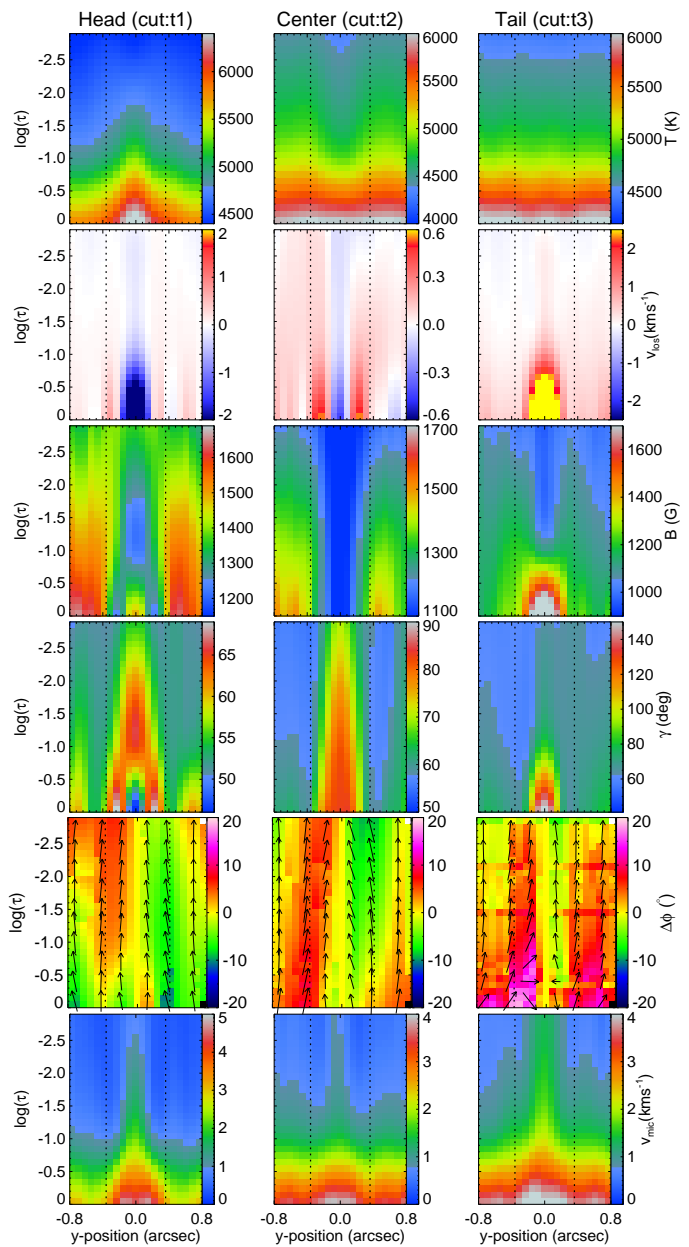}  
  \caption{Transverse-vertical slice of physical parameters of the standard filament.  For slice positions see \fig{fig:stand_fil}. Two vertical dotted lines represent the boundary of the filament. Any offset in the magnetic azimuth is removed, and overplotted vectors are for illustration purposes only.}
  \label{fig:slice_trans}
\end{figure}

All features seen in the position-dependent averages are even more clearly visible in the global average since the fluctuations around the average are further reduced. The height-dependent properties of the standard filament are highlighted in \fig{fig:slice_long}, which depicts a vertical cut along the axis of the filament, and in \fig{fig:slice_trans}, which shows three vertical cuts across the filament (t1, t2, and t3). The horizontal locations of the cuts are indicated in \fig{fig:stand_fil}. 

The temperature in the filament head displays a larger vertical gradient than in the rest of the filament, and reaches a value of $\approx$6600\,K at optical depth unity. Due to the dependence of the downwind thermal structure on the location of the filament within the penumbra, the temperature structure outside the head may be interpreted only qualitatively. Apart from the increase in temperature in the head and tail, a clear decrease in the temperature along the main axis of the filament can be observed in the highest layers of the atmosphere. This is particularly clearly visible in the top panel of \fig{fig:stand_fil} and in the transverse-height slice ``t2'', shown in \fig{fig:slice_trans}. This narrow, cool lane is probably what is observed as a dark lane above most of the filaments in the upper temperature map in \fig{fig:overview} and is not clearly visible in the middle photosphere. These dark lanes are best seen in the inner penumbra in our data, which may be a consequence of the difference in the thermal structure between filaments in the inner and outer parts of the penumbra. The continuum intensity structure of the standard filament is basically the same as the temperature structure at $log(\tau)=0$.

The magnetic field strength along the filament axis is quite uniform, with a value of roughly 1000\,G, and close to horizontal at all heights, with the exception of the head and the tail of the filaments. It is interesting that the length of the region with a nearly horizontal field decreases with height, exactly as expected for a very shallow $\bigcap$-shaped loop-like structure. In the head and the tail, however,the field inclination significantly differs from horizontal, assuming average values close to 150$^\circ$ in the tail and close to 40$^\circ$ in the head. The magnetic field strength systematically takes on its largest values in the tail, regularly reaching values of 1500--2500\,G in the deepest layers, while individual tails harbour nearly vertical fields with a strength of up to 3500\,G, \citep[but see also][]{downflowpaper}.

As can be seen from \fig{fig:slice_trans}, in both the head (cut ``t1'') and the tail (cut ``t3'') the field is comparatively vertical, but as one moves away from the axis of the filament the field becomes nearly horizontal (at the head this is true only in the lowest layer) before pointing upward at still greater distances (outside the filament). At the tail this horizontal field is simply the interface between the downward pointing field in the filament and upward pointing field outside it and is not associated with any special feature in velocity or field strength. However, around the head this systematically present horizontal field, about 0.2\arcsec{} from the axis, coincides with the lateral downflows and with a minimum in field strength. This combination is suggestive of cancellation of unresolved opposite polarity signal. This interpretation is supported by many of the individual filaments in all parts of the penumbra, where a reversal of the field direction can be observed directly at this location (see e.g. \fig{fig:incl_opp_pol}).

The upflow velocity in the head of the standard filament peaks at around 5 \kms{}, with the upflows in individual filaments reaching values up to 11 \kms{}, which is significantly higher than the estimated sound speed (5--8 \kms{}). The transverse-vertical slices of the line-of-sight velocity in \fig{fig:slice_trans} clearly show the narrow downflow channels on each side of the upflow in the filament's head, gradually widening towards the filament axis with distance along the filament, to eventually merge and form the strong downflow channel in the tail of the filament. The width of the filament remains approximately constant over the entire length and only reduces slightly towards the tail. The line-of-sight velocity in the lateral downflow channels has a typical value of $\sim$ 0.5 \kms{}, which quickly weakens with height and disappears at an optical depth of approximately 0.1. Downflow velocities in the tail are significantly higher, with an average value of 7 \kms{} over all filaments and a maximum of up to 19 \kms{} in individual filaments. This concentrated downflow is supersonic in nearly half of the filaments. 

The azimuth relative to the filament axis is clearly the noisiest quantity, especially in the top and bottom layers of the atmosphere. In addition, in the bottom two atmospheric layers an offset of several degrees possibly is an indication of a small polarimetric calibration error. To realize a clear picture, as described before in Section \ref{sec:select}, we have removed the offset in the azimuth such that it is zero at the central axis. A clear pattern is visible, with negative values dominating on the left of the filament, and positive values being present mainly on the right side of the filament, in particular in the upper layers, indicating that the magnetic field is converging toward the filament axis. This is illustrated by the arrows in \figs{fig:stand_fil}{fig:slice_trans}. We stress that the arrows are meant to only very qualitatively guide the eye, since due to the corrugated $\tau$ surfaces, they may well point in other directions in a picture in which $log(\tau)$ is replaced by a geometrical height $z$.   

The microturbulent velocity of the averaged penumbral filament shows a weak double peak in the transversal direction in the deep layers at the location of the head, as can be seen in \fig{fig:slice_trans} (lowest left panel). The location of the maxima coincides with the boundary between the central upflow in the head and the narrow downflow channels on the sides and indicates that the boundary between the counterstreaming flows is not resolved. This pattern persists in the downstream transversal slices, but is much weaker, possibly due to decreasing difference between the up- and downflow velocities. 

In summary, we find a clear picture of the internal structure of penumbral filaments, showing a complex pattern of flows and a well-defined magnetic and thermal structure. 

 
\subsection{Mass flux\label{sec:massflux}}

We estimate the vertical mass flux in the penumbral filaments by taking the product of the density $\rho$ and the line-of-sight velocity $v_{los}$ returned by the inversion code. This is inevitably a very crude estimate at best, since the SPINOR inversion code calculates the density using hydrostatic equilibrium, an approximation that is not valid in a strongly magnetized atmosphere. In addition, since the vertical coordinate relevant for the inversions is not a geometric coordinate, but indicates constant optical depth, the line-of-sight velocity is not necessarily normal to the surface over which we calculate the flux, because of the corrugated shape of the penumbral filaments, as inferred by \cite{zakh08}. In particular, since the opacity is a strong function of the temperature, the optical depth surfaces will be inclined towards higher layers in the direction of any temperature gradient. 

\begin{figure}[ht]
  \centering
  \includegraphics[width=\linewidth]{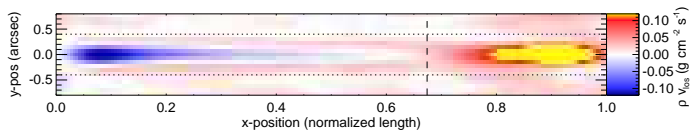}
  \caption{Mass flux ($\rho~ v_{los}$) of the standard filament.  The two horizontal dotted lines bound the filament on its sides.  The transversal dashed line represents the boundary of upflows along the central axis of the filament. Mass flux is computed for the region between the two horizontal dotted lines only.}
  \label{fig:slice_long_massflux}
\end{figure}

Despite these limitations, an estimate of the mass flux can help to get an impression of the important flows in the structure, in particular, which fraction of the mass that flows up in the head and along the axis of a filament flows back below the surface again in the tail (radial) and which fraction along the sides of the filament (azimuthal). To this end, we integrate the mass flux $\rho v_{los}$ over the area of the filament, defined to be the area enclosed by the two horizontal dotted lines in \fig{fig:slice_long_massflux}. 

The integration yields an excess in the downward flux of some three times the upward flux. Upflowing mass flux comes out to be -2$\times 10^{11}$ kg s$^{-1}$ and downflowing mass flux 6$\times 10^{11}$ kg s$^{-1}$, both computed at the log$(\tau)$=0 level for a filament of average length. The seeming excess of downflowing material may well be a consequence of the inclination of the optical depth surface, which can lead to a significant mass flux slipping  across in the horizontal direction unnoticed. 

\cite{west97} found the downflowing mass flux in the outer penumbra to be twice the upflowing mass flux in the inner penumbra of the sunspot they studied. Our result is in qualitative agreement with that obtained by \cite{west97}. However, both our study and that of \cite {west97} at first sight appear to disagree with the results obtained by \cite{pusc10a}, that the upward mass flux computed at log$(\tau)$=0 over a part of the inner penumbra of a sunspot is five times larger than the downflowing mass flux. This apparent inconsistency is most likely due to \cite{pusc10a} evaluating the mass flux over a limited field of view, situated in the inner penumbra, a region dominated by penumbral heads containing upflows. The fact that they concentrated on the disc-center side of their spot, thus projecting the Evershed flow as a blue-shift into the line of sight probably also contributed, as possibly did the fact that they considered the mass fluxes at a fixed geometrical height rather than a given optical depth.

To estimate the fraction of the mass that returns to the photosphere in the narrow downflow lanes at the side of the filament, we repeat the integration, but separately on the two sides of the vertical dashed line in \fig{fig:slice_long_massflux}. We obtain a downflowing mass flux of 1.5$\times 10^{11}$ kg s$^{-1}$ in the upstream part and 4.5$\times 10^{11}$ kg s$^{-1}$ in the downstream (tail) part. This indicates that at least 25\% of the total downflowing material returns to the solar interior in the lateral downflow lanes. 

Although the mass flux is determined less and less reliably with increasing height, because of the smaller velocities and lower reliability of the inversions in the highest layers, indications are that the fraction of mass flowing back to the surface in the first half of the filament decreases with height, suggesting that the mass flow in penumbral filaments becomes more radial with increasing height in the atmosphere.

\section{Discussion}
\label{sec:discussion}

The most striking result emerging from our investigation is the degree of uniformity between all the selected filaments in the penumbra. Although variations do exist between the individual filaments, they are more quantitative rather than qualitative in nature and do not hide the commonalities between the filaments. Thus, already averaging over only a small number of filaments results in a picture that is by and large representative of all locations in the sunspot penumbra. This study also highlights that filaments are one of the main constituents of the penumbra. Together with spines, intrusions of near-umbral material into the penumbra, they describe most of the penumbra's structure. Other penumbral fine structure is likely associated with these two types of features in some way, e.g. being part of them. Thus, penumbral grains \citep{mull73,rimm06} are probably nothing else than the tear-shaped bright heads of the filaments.

Another important point emerging from our study is that there are no `bright' and `dark' filaments, a concept that plays an important role in the earlier literature on the penumbra. Instead there are brighter and darker {\em parts} of each filament. As we discuss later (in Section \ref{sec:controversies}), this distinction can clarify a number of controversies that have raged in the literature without having been resolved \cite[for a detailed discussion, see][]{sola03}.
 
We divide this section into two subsections. In the first one, we discuss the structure of penumbral filaments based on the results obtained in the previous section, and the compatibility of different models and simulations with it. In the second subsection, we discuss some controversies regarding the true structure of spines/inter-spines (filaments) and the associated flows.

\subsection{Structure of penumbral filaments} 

In the past, the brightness of penumbral features has been extensively inferred from the continuum intensity images of sunspots \cite[for an overview, see][]{sola03,borr11,remp11}. We find that the continuum intensity exhibits a behaviour very similar to the temperature at $log(\tau)=0$, as expected from an LTE inversion. 

The universally hotter heads (with comparatively little variation from filament to filament) and cooler tails of penumbral filaments found here have not been reported before in this generality (but see, \citealt{borr05} and \citealt{bhar10}). Furthermore, the relatively common secondary peak in the temperature in the tail has previously not been noticed. The dependence of the temperature structure within a filament on the distance to the umbra has also not been described before and is consistent with the changes observed in the environment of the filaments. The temperature of the environment in which the filaments are embedded increases systematically from the inner to the outer penumbra. This restricts the ability of the outer filaments to cool efficiently, contributing to the flatter radial temperature profile of these filaments. This change in the environment reflects the decreasing width and number density of the spines towards the outer penumbra.

The dark cores of filaments, first detected by \cite{scha02} and later confirmed by \cite{bell05,bell07,lang07,rimm08}, are best seen in images of temperature in the middle and upper photospheric layers of inner penumbral filaments. They are also clearly visible in the standard filament as a reduction of the temperature by 200--400\,K in layers above $log(\tau)=-0.9$. They are associated with a weaker and more horizontal magnetic field than their surroundings, in agreement with the observations of \cite{bell07} and \cite{lang07}. Recently, the dark cores have been interpreted as a signature of field-free gaps, the suggestion being that the absence of magnetic pressure in the gap results in a higher gas pressure thereby raising the $\tau=1$ surface to higher and thus cooler layers, which then become visible as a dark lane \citep{lang07}. An equivalent explanation of the dark cores based on the concept of flux tubes is given by \cite{borr07} and \cite{ruiz08}. They suggested that these are a signature of increased density due to weaker fields in horizontal flux tubes, which shifts the optical depth unity surface towards higher and cooler layers. 

The temperature decrease at small $\tau$ along the bulk of the filament appears to be compatible with the idea of a raised $\tau$ = 1 level (e.g., due to an enhanced density). The field from the spines is expected to close over the filaments at heights above these cool layers. The larger number of spines with strong magnetic field in the inner penumbra may explain the greater preponderance of dark lanes in inner penumbral filaments, in particular near the filament's heads.    

However, the observed field strength of around 1\,kG present throughout the bulk of the filaments is relatively strong. This indicates that although the penumbral filaments have a somewhat lower field strength than their surroundings, particularly in the inner penumbra (cf. \citealt{borr06}), they are not field-free. This agrees well with the magnetized character of the Evershed flow, as observed by \cite{sola94,borr05,ichi08a} and \cite{borr10}.

Strong upflows in the inner parts of penumbrae and strong downflows in their outer parts are a well established property of sunspot penumbrae and have been reported by many researchers \citep[see, e.g.,][]{schl00,west01a,trit04,sanc07,ichi07,fran09,downflowpaper}. How this global pattern is related to the velocity structure in individual penumbral filaments has remained unresolved, however. We find that penumbral filaments contain strong upflows ($\sim 5$ \kms{} on average) in their heads and even stronger downflows ($\sim 7$ \kms{}) in their tails, similar to the flow speeds reported earlier in the inner and outer parts of sunspot penumbrae. The maximum values of the up- and downflows in the individual filaments can be considerably higher than these averages, at $\sim 11$ \kms{} and $\sim 19$ \kms{}, respectively. Around 20\% of the heads and more than 50\% of the tails harbor flows whose speed exceeds the sound speed of $\sim$7 \kms{}. In addition to these concentrated rapid flows we also observed upflows extending along the axis of each filament and downflows along narrow lanes at the lateral edges of the filament.  
Since no significant up- or downflows are seen outside the filaments, these up- and downflows must be connected by a horizontal flow which is not visible in our data because of the vertical line-of-sight. The component of this flow along the filament axis is the Evershed flow. 

Although supersonic downflows in the middle or outer parts of the penumbra of different sunspots were observed by a number of researchers \citep[see, e.g.,][]{del01,bell04,pill09}, these  downflows were observed either in very small areas (single pixels: e.g., \citealt{del01,pill09}), hiding their internal structure, or in much larger regions than we find (e.g., \citealt{bell04}), suggestive of a different origin. The partly even faster downflows found by \cite{downflowpaper} have some commonalities with the rapid downflows considered here (thus, both are located at the ends of filaments), but the fastest ones from that study also show clear differences, being larger and located mainly at the ends of complex filaments having multiple heads. Often they are also larger than the downflows at the ends of normal filaments.   

Contrary to the claims of \cite{fran09,bell10} and \cite{pusc10a}, the presence of downflows in the inner penumbra has been reported by \cite{sanc07,josh11,scha11,scha12}\footnote{The downflows seen at the inner edge of the penumbra by \cite{schl99a} are located in very cool gas, which appears to be in umbral intrusions. These downflows may be spurious, caused by a blend to the \CI\ 5380\AA\ line at umbral temperatures \citep{scha11,josh11}.}. In a recent study, \cite{scha13} report narrow downflow lanes to be located at the boundaries between spines and inter-spines. Here we extend this result by showing that the downflows in a filament form a V shape, with the tail of the filament coinciding with the bottom of the V, and the head lying between the two arms at the top. 

The picture of the horizontal flows in the standard filament (deduced from the line-of-sight velocities) is one of a smooth transition between a radial and an azimuthal flow, with the azimuthal part dominating the upwind half and the radial part the downwind half of the filament. 
Furthermore, the clear, quantitative relationship between the direction of flow and the temperature, with upflowing material being hotter than the downflowing material (the lower part of each panel in \fig{fig:scat}), provides to our knowledge the clearest support for the presence of overturning convection in penumbral filaments so far. The departure from this behavior seen in the upper half of each panel in \fig{fig:scat} is seen only in the strong downflow in the tail and may be caused by the strong magnetic field found there, lowering the optical depth unity surface to deeper and hotter layers.

The magnetic field in filaments has a well-defined 3-part structure: A strong (1.5-2kG) nearly vertical field (with same polarity as umbra) at the head of the filament, a kG nearly horizontal field pointing mainly from the head to the tail in the body of the filament and a nearly vertical, with opposite polarity to umbra, strong (2-3.5kG) field at the filament's tail. The strength of the field in the tail increases with the downflow speed there, as found by \cite{downflowpaper} and can, for the most extreme flows studied by them, even exceed umbral values without being associated with umbral darkness. 



The boundary between the magnetic field of the penumbral filaments and that of the surrounding material is clearly marked by a sharp change, in particular in the magnetic field inclination. Although the relatively vertical lanes outside the filaments on both sides of the filament's head in Fig.~\ref{fig:slice_long} and \ref{fig:slice_trans} (i.e., in location ``t1''), suggest that a separation of the penumbra into ``spines'' i.e., regions of strong, relatively vertical field, and ``inter-spines'', regions with relatively inclined field, is justified, the inclination by itself is not sufficient to distinguish between the nearly vertical field in the filament's heads and the similarly inclined field in the spines. Moreover, it is clear from Fig.~\ref{fig:trans_cut} that, in contrast to the filaments, the inclination of their surroundings varies significantly across the penumbra, with the field in the surroundings being weaker and more horizontal in the outer than in the inner penumbra. 
There is a noticeably larger variation with distance from the umbra in the inclination and field strength of the surroundings of the filaments than of the filaments themselves. This is not surprising, since the tracking procedure was focused on the selection of filaments only.
Nonetheless, we expect that this is partly also because the magnetic field in spines weakens and becomes more horizontal with distance from the umbra. Partly, however, this is also because the spines are thinner and less common in the outer penumbra, where filaments tend to border on other filaments.

The behaviour of the magnetic vector's azimuth angle across the filament in the deepest layers is noisy and not consistent from filament to filament, e.g., at different radial distances in the penumbra. However, averaging over all sixty selected filaments reveals that the field in the deepest layers of the filament head is diverging slightly away from the filament axis towards the narrow downflow channels at the sides of the filament near its head.
The magnetic field has a slightly lower strength at the location of these downflow channels. This is accompanied by an increase in the inclination angle, with the field changing sign and pointing downwards in these lateral downflow lanes in many of the selected filaments (see e.g. \fig{fig:incl_opp_pol}), as reported recently by \cite{ruiz13} and \cite{scha13}. This behavior suggests that there is opposite polarity field located near the lateral boundary of the filament, right next to the magnetic field of the ``spines'', although, due to cancellation of its Stokes V signal with that from the upward directed field in the neighbouring spines, the opposite polarity is sometimes not seen (e.g., in the averaged filaments, where any variation in the width of the individual filaments being averaged will smear out narrow signals). The downward directed field on the sides of the filament probably stems from the horizontal field in the filament that is pulled down by downflowing material in the lanes located there, as described by \cite{remp12}.


The field divergence seen in the deeper layers gives way to a field convergence towards the filament's central axis in the upper layers, similar to the findings of \cite{borr08}. The converging field extends outside the filaments and is probably rooted there, suggesting that this field originates in the neighbouring spines and expands over the filaments, which are restricted to the lower photosphere (a geometry first proposed by \citealt{sola93}).  

It is unclear what happens to the magnetic field that submerges at the ends of the filaments. This field may remain below the solar surface until well outside the sunspot as proposed by \cite{thom02}, or instead may form a structure similar to a $\bigcup$-loop returning to the surface within the penumbra \citep{schl10} and possibly moving outwards \citep{sain08}, appearing as a pair of moving magnetic features outside the penumbra as proposed by \cite{zhan03}. Reality may be more complex, however, with the field possibly dispersing below the solar surface. 

To propel the Evershed flow, two mechanisms have been proposed, namely a (siphon) flow driven by a magnetic field gradient \citep{meye68,thom93,dege93,mont97}, and a thermally driven or convective flow \citep{schl98a,schl98b,remp11a}. \cite{scha08} and \cite{remp09b}, using MHD simulation results of penumbral filaments (or proto filaments in the case of \citealt{scha08}) in a simplified configuration, show that the inclined magnetic field in a filament diverts the convective upflow into a horizontal flow, directed mainly ``radially'' (i.e. along the filament axis), that they associate with the Evershed flow. 

Our results clearly show a systematic temperature decrease of 400-800 K from the head to the tail of the filament, which is compatible with the convective driver scenario. In this scenario the convective instability (modified for the presence of a magnetic field) accelerates both, hot up- and cool downflows, below the solar surface. Once the upflowing gas reaches the (convectively stable) solar surface layers, it decelerates, building up excess pressure there. This accelerates the rising gas sideways. Due to the general radially outward inclination of the magnetic field the gas flows to a large extent radially outward, i.e. towards the tail of the filament. However, the horizontal velocity generally also has an azimuthal component, i.e. directed from the axis to the sides of the filament. 
The interpretation of the Evershed flow as the radial component of a convective flow is compatible with the observed flow pattern of the penumbral filaments, which have some similarity with highly skewed and elongated convective cells. Due to the strong dependence of the opacity on the temperature we see higher layers in the head than in the tail, so that the temperature difference at equal geometrical height should be considerably larger \citep{brul99}. The pressure difference associated with this horizontal temperature gradient may contribute to the acceleration of the Evershed flow, but to determine this a geometric height scale is needed. 



The possibility that a siphon flow mechanism drives the Evershed flow cannot be ruled out, however. We find a systematic excess magnetic field strength of some 500\ G in the tail over that in the head of the filaments, which may contribute to the acceleration of the flow, since the enhanced field strength in the tail implies that the gas pressure is larger in the head than in the tail, in accordance with the model of \cite{meye68}, cf. e.g., \cite{mont97}. However, to confirm this a geometrical scale of the parameters will be necessary. Thus, \cite{downflowpaper} showed that in the strongest downflows, associated with the strongest fields, in penumbrae the optical depth unity surface is strongly depressed.

The lateral flow pattern observed in the deepest layers shows similarities to the convective flows in the field-free gap model of \cite{scha06} and \cite{spru06}. However, the observations indicate that real filaments are more complex than this model suggests. The strongly magnetized nature of the Evershed flow channel and the localized strong upflows/downflows near the heads/tails of the filaments are not described by this model.

The flux tube models \citep{sola93,schl98a,schl98b,borr05} are supported by the measured strength and inclination of the magnetic field and can, for a flux tube emerging and submerging again, explain the localized strong upflow/downflow regions near heads/tails of the filaments, although the local strong changes in the inclination are rather unexpected. Such models can also reproduce the tear-shaped bright heads of filaments \citep{schl99}, but have difficulty in explaining the observed upflows along more than half of the length of the filaments and downflows containing downward pointing magnetic field along the sides. In particular, lateral convection cannot be represented in a model based on thin flux tubes. 

The picture emerging from our observations is a complex one that combines aspects of both the models and bears a strong resemblance to the results of numerical simulations \citep{remp09b,remp12}, although some quantitative differences appear to be present.   

\subsection{Long-standing controversies\label{sec:controversies}} 

Our investigation reveals that spines and (parts of) filaments have some properties in common. For example, both of them harbour vertical, umbral polarity fields (throughout the spines, but restricted to the heads of filaments). This can lead to confusion if a height-independent magnetic inclination alone is used to distinguish between spines and inter-spines. One major difference between the filaments and spines is that the filaments are strongly height dependent and are best seen in the very deepest layers of the atmosphere. In the higher layers the expanding field of the spines closes over the interposing filaments.  

A major source of confusion is that the properties of filaments change significantly along their length. Their temperature and brightness in their downstream parts can assume values that are closer to the temperature of spines than to that of filament heads. Such changes along the bodies of filaments can easily lead to seemingly contradicting correlations between different physical parameters, such as magnetic field strength, inclination, brightness, and velocity of insufficiently resolved penumbral filaments.


For example, the more horizontal fields were found in darker penumbral regions by \cite{beck69,schm92,stan97} and \cite{wieh00}, whereas \cite{lites93} found the darker parts of the penumbra to host the more vertical field, although the correlation remained non-conclusive; see also \cite{skum94} for a critical review of the correlation between inclination and brightness of penumbral structures. Whereas \cite{lites93} compared spines with inter-spines, we assume that the other authors compared the horizontal fields in the dark filament bodies with the more vertical ones in the bright filament heads. The stronger field strength in the filament's head, tail and surroundings (spines) results in the absence of a general relationship between the brightness and the magnetic field strength in sunspot penumbrae, as concluded by \cite{wieh89} and \cite{schm92}. 

Similarly, the inferences by \cite{west01a} and \cite{lang05} that the spines are warmer, may represent a misperception caused by the more vertical field at the bright heads of penumbral filaments, which were then mistaken for spines. The findings of \cite{beck69} and \cite{wieh00} that dark penumbral features are associated with stronger (and more horizontal) magnetic fields may be the result of signal cancellation between unresolved spines and tails of penumbral filaments, compounded by missing linear polarization data.
A recent conclusion drawn by \cite{borr11}, that the inter-spines are brighter filaments in the inner penumbra and darker filaments in the outer penumbra seems to be a result of looking at different parts of filaments in the inner and outer penumbra. Our results actually reveal the opposite behaviour, filaments in the outer penumbra have a somewhat warmer body and tail than their counterparts in the inner penumbra. 

Another such controversy surrounds the question whether the Evershed flow mainly takes place in dark penumbral filaments, as has been argued by, e.g., \cite{titl93} and \cite{west01a}, or if no correlation between flow and brightness exists \citep[e.g.,][]{wieh89,lites90,hirz01}. In addition to the explanations proposed in the literature (e.g., \citealt{dege94}; see a more complete discussion and many more references in the review by \citealt{sola03}), the complex structure of the individual filaments allows for both points of view, depending on the spatial resolution, on the part of the penumbra considered, on the spectral line, etc. We intend to consider this controversy more closely with the help of a sunspot observed closer to the limb, which displays the Evershed flow clearly.  

The universal structure of penumbral filaments derived here is able to resolve the above, partly long-standing misconceptions and controversies.
This discussion demonstrates the importance of the concept of complex, but nearly universal filaments for understanding the observations of sunspot penumbrae.


\section{Conclusions}
\label{sec:conclusion}
We have studied the structure of penumbral filaments by applying a newly developed spatially coupled inversion technique to a sunspot observed by Hinode (SOT/SP) almost exactly at the solar disc center. The most striking result of this inversion is that all the simple penumbral filaments (i.e., those with a single head and a single tail) are rather similar to each other, irrespective of their position in the penumbra, barring some quantitative differences in the temperature. These similarities do not just set constraints on the models  and simulations of penumbral convection, they also justified the creation of an averaged, ``standard'' filament. Our high-resolution study of filaments clarifies a number of features of penumbral fine-scale structure. For example, penumbral grains can be identified with the drop-shaped hot and bright heads of filaments. Hence, the sunspot penumbra is found to be composed mainly of filaments and spines. Earlier classifications of the penumbra into bright and dark filaments are not supported by our work. Instead, we find that each filament has a bright and a dark part.

Filaments are found to have a characteristic magnetic structure. A significant, approximately horizontal magnetic field of $\sim$ 1000\,G is found throughout the main body of the filaments. Whereas an enhanced (1.5--2\,kG) vertical field with the polarity of the umbra is present in the heads, and a strong (2--3.5\,kG) vertical magnetic field of opposite polarity is observed in the tails. An opposite polarity field is also observed along the sides of the filaments. 

We have presented evidence for the presence of magneto-convection in the penumbral filaments, in that the vertical flows in a filament give the impression of a strongly elongated convection cell (like a strongly stretched granule), with upflows dominating at the filament's head and downflows at its tail, but upflows also being present along the central axis of the filament and narrow downflow lanes along its sides. This suggests that the overturning convection takes place, both in the lateral (i.e., azimuthal) and radial directions. 
Importantly for this interpretation, there is a clear, quantitative relationship between flow velocity and temperature. Thus, we find upflows to be warmer than downflows by up to 800K at the $log(\tau)$=0 layer. Due to the strong dependence of the $log(\tau)$=0 level on the temperature, we expect the temperature difference at a fixed geometrical height to be significantly larger. The strongest downflows (in the tails of the filaments) are found to be 50-150K warmer than most other downflows, suggesting a local source of heating \citep[see][for a detailed study of such rapid downflows]{downflowpaper}. 

From the structure of the field and the line-of-sight flows we conclude that the upflow feeding the Evershed flow is mainly concentrated in the head of a penumbral filament, which is brighter and associated with an upward pointing, strong field. The gas caught in the Evershed flow cools while it travels along the filament's bulk following the nearly horizontal field. This gas sinks as comparatively cool and dark material at the tail of the filament associated with a strong downward pointing field. It may get heated again by a shock formed when the supersonically sinking gas shows to subsonic speed. A part of the gas emerging along the filament's axis flows sideways, dragging and twisting the magnetic field on its way. This gas flows down in narrow lanes adorning the sides of the filament. There the field points downwards. Several long-standing controversies on the Evershed flow being concentrated in bright vs. dark penumbral structures, and it being associated with strong vs. weak fields can potentially be resolved by the observed changes in the brightness and magnetic field along the length of a filament. Observations that do not resolve filaments and do not distinguish all their parts clearly from spines can easily lead to contradicting results. 



An approximate evaluation of the mass flux balance reveals a strong (factor of $\approx$3) excess of downflowing mass compared with upflows. This suggests that a significant amount of the material flowing up in the upstream part of the filament emerges unseen, probably due to the strongly corrugated iso-tau surface, which lies higher in the hot upflows than in the cool downflows, so that mass can flow through this surface horizontally without becoming visible in the line-of-sight velocity at solar disc centre. 

Although the temperature of the heads of the filaments is independent of the position in the penumbra, the thermal profile of the filaments depends on the distance of the filament to the umbra. This may be due to a decreased ability to cool by lateral radiative losses in the warmer environment of the outer penumbra as compared to the inner penumbra, where more cool spines are present. 

Dark cores \citep{scha02} are seen as a decrease in the temperature by 200--400\,K in the filament's middle and upper photospheric layers, which is compatible with the height at which the expanding magnetic field from the neighbouring spines merges over the filament.

The structure of the penumbral filaments as a whole bears aspects of both the flux tube and field-free gap models, and shows strong qualitative similarities with the model described by \cite{remp12}. A quantitative comparison with MHD simulations of sunspots will be the topic of a separate publication.

Long-standing disagreements on the spines/inter-spines being bright/dark penumbral structures, and being associated with vertical/inclined, strong/weak magnetic field, are easily resolved in the light of the complex structure of individual filaments, which exhibit both bright and dark parts, nearly vertical as well as nearly horizontal fields, etc. E.g., strong, relatively vertical fields are found both in the bright heads of the filaments (or inter-spines) and in the dark spines.

Although various aspects of the convective structure of the penumbral filaments have been reported in a number of publications in the past, we can for the first time give a relatively complete description that is unified in two senses: it describes together all the salient properties of filaments (with the exception of those related to their evolution) and it is applicable to all simple filaments in a sunspot. 

The main limitation of the present study is that it does not provide any information on the evolution of penumbral filaments with time. Obtaining such information will be one of the aims of future flights of the Sunrise balloon-borne observatory \citep{sola10,bart11}. 

We also plan to follow up on this work by investigating the penumbral fine structure of a sunspot observed closer to the solar limb, where the Evershed flow can be directly seen. 


\begin{acknowledgements}
Hinode is a Japanese mission developed and launched by ISAS/JAXA,
collaborating with NAOJ as a domestic partner, NASA and STFC (UK) as
international partners. Scientific operation of the Hinode mission is
conducted by the Hinode science team organized at ISAS/JAXA.
This team mainly consists of scientists from institutes in the partner countries.
Support for the post-launch operation is provided by JAXA and NAOJ (Japan),
STFC (U.K.), NASA (U.S.A.), ESA, and NSC (Norway). This work has been partially 
supported by the WCU grant number R31-10016 funded by the Korean Ministry of 
Education, Science and Technology. This research has made use of NASA's Astrophysics Data System. 
\end{acknowledgements}


\begin{thebibliography}{125}
\expandafter\ifx\csname natexlab\endcsname\relax\def\natexlab#1{#1}\fi

\bibitem[{{Barthol} {et~al.}(2011){Barthol}, {Gandorfer}, {Solanki},
  {Sch{\"u}ssler}, {Chares}, {Curdt}, {Deutsch}, {Feller}, {Germerott},
  {Grauf}, {Heerlein}, {Hirzberger}, {Kolleck}, {Meller}, {M{\"u}ller},
  {Riethm{\"u}ller}, {Tomasch}, {Kn{\"o}lker}, {Lites}, {Card}, {Elmore},
  {Fox}, {Lecinski}, {Nelson}, {Summers}, {Watt}, {Mart{\'{\i}}nez Pillet},
  {Bonet}, {Schmidt}, {Berkefeld}, {Title}, {Domingo}, {Gasent Blesa}, {Del
  Toro Iniesta}, {L{\'o}pez Jim{\'e}nez}, {{\'A}lvarez-Herrero},
  {Sabau-Graziati}, {Widani}, {Haberler}, {H{\"a}rtel}, {Kampf}, {Levin},
  {P{\'e}rez Grande}, {Sanz-Andr{\'e}s}, \& {Schmidt}}]{bart11}
{Barthol}, P., {Gandorfer}, A., {Solanki}, S.~K., {et~al.} 2011, \solphys, 268, 1


\bibitem[{{Beckers} \& {Schr{\"o}ter}(1969)}]{beck69}
{Beckers}, J.~M. \& {Schr{\"o}ter}, E.~H. 1969, \solphys, 10, 384

\bibitem[{{Bellot Rubio}(2010)}]{bell10a}
{Bellot Rubio}, L.~R. 2010, in Magnetic Coupling between the Interior and
  Atmosphere of the Sun, ed. S.~S. {Hasan} \& R.~J. {Rutten}, 193

\bibitem[{{Bellot Rubio} {et~al.}(2004){Bellot Rubio}, {Balthasar}, \&
  {Collados}}]{bell04}
{Bellot Rubio}, L.~R., {Balthasar}, H., \& {Collados}, M. 2004, \aap, 427, 319

\bibitem[{{Bellot Rubio} {et~al.}(2005){Bellot Rubio}, {Langhans}, \&
  {Schlichenmaier}}]{bell05}
{Bellot Rubio}, L.~R., {Langhans}, K., \& {Schlichenmaier}, R. 2005, \aap, 443,
  L7

\bibitem[{{Bellot Rubio} {et~al.}(2007){Bellot Rubio}, {Tsuneta}, {Ichimoto},
  {Katsukawa}, {Lites}, {Nagata}, {Shimizu}, {Shine}, {Suematsu}, {Tarbell},
  {Title}, \& {del Toro Iniesta}}]{bell07}
{Bellot Rubio}, L.~R., {Tsuneta}, S., {Ichimoto}, K., {et~al.} 2007, \apjl,
  668, L91

\bibitem[{{Bellot Rubio} {et~al.}(2010){Bellot Rubio}, {Schlichenmaier}, \&
  {Langhans}}]{bell10}
{Bellot Rubio}, L.~R., {Schlichenmaier}, R., \& {Langhans}, K. 2010, \apj, 725,
  11

\bibitem[{{Bharti} {et~al.}(2012){Bharti}, {Cameron}, {Rempel}, {Hirzberger},
  \& {Solanki}}]{bhar12}
{Bharti}, L., {Cameron}, R.~H., {Rempel}, M., {Hirzberger}, J., \& {Solanki},
  S.~K. 2012, \apj, 752, 128

\bibitem[{{Bharti} {et~al.}(2010){Bharti}, {Solanki}, \& {Hirzberger}}]{bhar10}
{Bharti}, L., {Solanki}, S.~K., \& {Hirzberger}, J. 2010, \apjl, 722, L194


\bibitem[{{Borrero}(2007)}]{borr07}
{Borrero}, J.~M. 2007, \aap, 471, 967

\bibitem[{{Borrero}(2009)}]{borr09}
{Borrero}, J.~M. 2009, Science in China G: Physics and Astronomy, 52, 1670

\bibitem[{{Borrero} \& {Ichimoto}(2011)}]{borr11}
{Borrero}, J.~M. \& {Ichimoto}, K. 2011, Living Reviews in Solar Physics, 8, 4

\bibitem[{{Borrero} {et~al.}(2005){Borrero}, {Lagg}, {Solanki}, \&
  {Collados}}]{borr05}
{Borrero}, J.~M., {Lagg}, A., {Solanki}, S.~K., \& {Collados}, M. 2005, \aap,
  436, 333

\bibitem[{{Borrero} {et~al.}(2008){Borrero}, {Lites}, \& {Solanki}}]{borr08}
{Borrero}, J.~M., {Lites}, B.~W., \& {Solanki}, S.~K. 2008, \aap, 481, L13

\bibitem[{{Borrero} \& {Solanki}(2010)}]{borr10}
{Borrero}, J.~M. \& {Solanki}, S.~K. 2010, \apj, 709, 349

\bibitem[{{Borrero} {et~al.}(2006){Borrero}, {Solanki}, {Lagg},
  {Socas-Navarro}, \& {Lites}}]{borr06}
{Borrero}, J.~M., {Solanki}, S.~K., {Lagg}, A., {Socas-Navarro}, H., \&
  {Lites}, B. 2006, \aap, 450, 383

\bibitem[{{Brandt} \& {Solanki}(1990)}]{bran90}
{Brandt}, P.~N. \& {Solanki}, S.~K. 1990, \aap, 231, 221

\bibitem[{{Brault} \& {Neckel}(1987)}]{brau87}
{Brault}, J. \& {Neckel}, H. 1987, Spectral Atlas of Solar Absolute Disk-Averaged and Disk-Center Intensity from 3290 to
12510 \AA\, unpublished

\bibitem[{{Bruls} {et~al.}(1999){Bruls}, {Vollm{\"o}ller}, \&
  {Sch{\"u}ssler}}]{brul99}
{Bruls}, J.~H.~M.~J., {Vollm{\"o}ller}, P., \& {Sch{\"u}ssler}, M. 1999, \aap,
  348, 233

\bibitem[{{Choudhuri}(1986)}]{chou86}
{Choudhuri}, A.~R. 1986, \apj, 302, 809

\bibitem[{{Cowling}(1953)}]{cowl53}
{Cowling}, T.~G. 1953, in The Sun, ed. G.~P. {Kuiper} (The University of
  Chicago Press), 532

\bibitem[{{Danielson}(1961{\natexlab{a}})}]{dani61a}
{Danielson}, R.~E. 1961{\natexlab{a}}, \apj, 134, 275

\bibitem[{{Danielson}(1961{\natexlab{b}})}]{dani61b}
{Danielson}, R.~E. 1961{\natexlab{b}}, \apj, 134, 289

\bibitem[{{de la Cruz Rodr{\'{\i}}guez} {et~al.}(2011){de la Cruz
  Rodr{\'{\i}}guez}, {Kiselman}, \& {Carlsson}}]{rodr11}
{de la Cruz Rodr{\'{\i}}guez}, J., {Kiselman}, D., \& {Carlsson}, M. 2011,
  \aap, 528, A113

\bibitem[{{Degenhardt} \& {Wiehr}(1991)}]{dege91}
{Degenhardt}, D. \& {Wiehr}, E. 1991, \aap, 252, 821

\bibitem[Degenhardt \& Wiehr(1994)]{dege94} 
 Degenhardt, D., \& Wiehr, E.\ 1994, \aap, 287, 620 

\bibitem[{{Degenhardt} {et~al.}(1993){Degenhardt}, {Solanki}, {Montesinos}, \&
  {Thomas}}]{dege93}
{Degenhardt}, D., {Solanki}, S.~K., {Montesinos}, B., \& {Thomas}, J.~H. 1993,
  \aap, 279, L29


\bibitem[{{del Toro Iniesta} {et~al.}(2001){del Toro Iniesta}, {Bellot Rubio},
  \& {Collados}}]{del01}
{del Toro Iniesta}, J.~C., {Bellot Rubio}, L.~R., \& {Collados}, M. 2001,
  \apjl, 549, L139


\bibitem[{{Evershed}(1909)}]{ever09}
{Evershed}, J. 1909, \mnras, 69, 454

\bibitem[{{Franz}(2011)}]{fran11}
{Franz}, M. 2011, PhD thesis, ArXiv e-prints:1107.2586

\bibitem[{{Franz} \& {Schlichenmaier}(2009)}]{fran09}
{Franz}, M. \& {Schlichenmaier}, R. 2009, \aap, 508, 1453

\bibitem[{{Frutiger} {et~al.}(2000){Frutiger}, {Solanki}, {Fligge}, \&
  {Bruls}}]{frut00}
{Frutiger}, C., {Solanki}, S.~K., {Fligge}, M., \& {Bruls}, J.~H.~M.~J. 2000,
  \aap, 358, 1109

\bibitem[{{Hirzberger} \& {Kneer}(2001)}]{hirz01}
{Hirzberger}, J. \& {Kneer}, F. 2001, \aap, 378, 1078

\bibitem[{{Ichimoto} {et~al.}(2007{\natexlab{a}}){Ichimoto}, {Shine}, {Lites},
  {Kubo}, {Shimizu}, {Suematsu}, {Tsuneta}, {Katsukawa}, {Tarbell}, {Title},
  {Nagata}, {Yokoyama}, \& {Shimojo}}]{ichi07}
{Ichimoto}, K., {Shine}, R.~A., {Lites}, B., {et~al.} 2007{\natexlab{a}},
  \pasj, 59, 593

\bibitem[{{Ichimoto} {et~al.}(2007{\natexlab{b}}){Ichimoto}, {Suematsu},
  {Tsuneta}, {Katsukawa}, {Shimizu}, {Shine}, {Tarbell}, {Title}, {Lites},
  {Kubo}, \& {Nagata}}]{ichi07a}
{Ichimoto}, K., {Suematsu}, Y., {Tsuneta}, S., {et~al.} 2007{\natexlab{b}},
  Science, 318, 1597

\bibitem[{{Ichimoto} {et~al.}(2008{\natexlab{a}}){Ichimoto}, {Lites}, {Elmore},
  {Suematsu}, {Tsuneta}, {Katsukawa}, {Shimizu}, {Shine}, {Tarbell}, {Title},
  {Kiyohara}, {Shinoda}, {Card}, {Lecinski}, {Streander}, {Nakagiri},
  {Miyashita}, {Noguchi}, {Hoffmann}, \& {Cruz}}]{ichi08}
{Ichimoto}, K., {Lites}, B., {Elmore}, D., {et~al.} 2008{\natexlab{a}},
  \solphys, 249, 233

\bibitem[{{Ichimoto} {et~al.}(2008{\natexlab{b}}){Ichimoto}, {Tsuneta},
  {Suematsu}, {Katsukawa}, {Shimizu}, {Lites}, {Kubo}, {Tarbell}, {Shine},
  {Title}, \& {Nagata}}]{ichi08a}
{Ichimoto}, K., {Tsuneta}, S., {Suematsu}, Y., {et~al.} 2008{\natexlab{b}},
  \aap, 481, L9

\bibitem[{{Jahn} \& {Schmidt}(1994)}]{jahn94}
{Jahn}, K. \& {Schmidt}, H.~U. 1994, \aap, 290, 295

\bibitem[{{Joshi} {et~al.}(2011){Joshi}, {Pietarila}, {Hirzberger}, {Solanki},
  {Aznar Cuadrado}, \& {Merenda}}]{josh11}
{Joshi}, J., {Pietarila}, A., {Hirzberger}, J., {et~al.} 2011, \apjl, 734, L18

\bibitem[{{Katsukawa} \& {Jur{\v c}{\'a}k}(2010)}]{kats10}
{Katsukawa}, Y. \& {Jur{\v c}{\'a}k}, J. 2010, \aap, 524, A20

\bibitem[{{Kosugi} {et~al.}(2007){Kosugi}, {Matsuzaki}, {Sakao}, {Shimizu},
  {Sone}, {Tachikawa}, {Hashimoto}, {Minesugi}, {Ohnishi}, {Yamada}, {Tsuneta},
  {Hara}, {Ichimoto}, {Suematsu}, {Shimojo}, {Watanabe}, {Shimada}, {Davis},
  {Hill}, {Owens}, {Title}, {Culhane}, {Harra}, {Doschek}, \& {Golub}}]{kosu07}
{Kosugi}, T., {Matsuzaki}, K., {Sakao}, T., {et~al.} 2007, \solphys, 243, 3

\bibitem[{{Langhans} {et~al.}(2005){Langhans}, {Scharmer}, {Kiselman},
  {L{\"o}fdahl}, \& {Berger}}]{lang05}
{Langhans}, K., {Scharmer}, G.~B., {Kiselman}, D., {L{\"o}fdahl}, M.~G., \&
  {Berger}, T.~E. 2005, \aap, 436, 1087

\bibitem[{{Langhans} {et~al.}(2007){Langhans}, {Scharmer}, {Kiselman}, \&
  {L{\"o}fdahl}}]{lang07}
{Langhans}, K., {Scharmer}, G.~B., {Kiselman}, D., \& {L{\"o}fdahl}, M.~G.
  2007, \aap, 464, 763

\bibitem[{{Leka} {et~al.}(2009){Leka}, {Barnes}, {Crouch}, {Metcalf}, {Gary},
  {Jing}, \& {Liu}}]{leka09}
{Leka}, K.~D., {Barnes}, G., {Crouch}, A.~D., {et~al.} 2009, \solphys, 260, 83

\bibitem[{{Lites} {et~al.}(1990){Lites}, {Skumanich}, \& {Scharmer}}]{lites90}
{Lites}, B.~W., {Skumanich}, A., \& {Scharmer}, G.~B. 1990, \apj, 355, 329

\bibitem[{{Lites} {et~al.}(1993){Lites}, {Elmore}, {Seagraves}, \&
  {Skumanich}}]{lites93}
{Lites}, B.~W., {Elmore}, D.~F., {Seagraves}, P., \& {Skumanich}, A.~P. 1993,
  \apj, 418, 928

\bibitem[{{Lites} \& {Ichimoto}(2013)}]{lites13}
{Lites}, B.~W. \& {Ichimoto}, K. 2013, \solphys, 283, 601

\bibitem[{{Mart{\'{\i}}nez Pillet} {et~al.}(2009){Mart{\'{\i}}nez Pillet},
  {Katsukawa}, {Puschmann}, \& {Ruiz Cobo}}]{pill09}
{Mart{\'{\i}}nez Pillet}, V., {Katsukawa}, Y., {Puschmann}, K.~G., \& {Ruiz
  Cobo}, B. 2009, \apjl, 701, L79



\bibitem[{{Metcalf}(1994)}]{metc94}
{Metcalf}, T.~R. 1994, \solphys, 155, 235

\bibitem[{{Meyer} \& {Schmidt}(1968)}]{meye68}
{Meyer}, F. \& {Schmidt}, H.~U. 1968, Mitteilungen der Astronomischen
  Gesellschaft Hamburg, 25, 194

\bibitem[{{Meyer} {et~al.}(1974){Meyer}, {Schmidt}, {Wilson}, \&
  {Weiss}}]{meye74}
{Meyer}, F., {Schmidt}, H.~U., {Wilson}, P.~R., \& {Weiss}, N.~O. 1974, \mnras,
  169, 35

\bibitem[{{Montesinos} \& {Thomas}(1993)}]{mont93}
{Montesinos}, B. \& {Thomas}, J.~H. 1993, \apj, 402, 314

\bibitem[{{Montesinos} \& {Thomas}(1997)}]{mont97}
{Montesinos}, B. \& {Thomas}, J.~H. 1997, \nat, 390, 485

\bibitem[{{Moore}(1981)}]{moor81}
{Moore}, R.~L. 1981, \apj, 249, 390

\bibitem[{{Muller}(1973)}]{mull73}
{Muller}, R. 1973, \solphys, 29, 55

\bibitem[{{Parker}(1979{\natexlab{a}})}]{parker79b}
{Parker}, E.~N. 1979{\natexlab{a}}, \apj, 230, 914

\bibitem[{{Parker}(1979{\natexlab{b}})}]{parker79a}
{Parker}, E.~N. 1979{\natexlab{b}}, \apj, 234, 333


\bibitem[{{Puschmann} {et~al.}(2010{\natexlab{a}}){Puschmann}, {Ruiz Cobo}, \&
  {Mart{\'{\i}}nez Pillet}}]{pusc10a}
{Puschmann}, K.~G., {Ruiz Cobo}, B., \& {Mart{\'{\i}}nez Pillet}, V.
  2010{\natexlab{a}}, \apj, 720, 1417

\bibitem[{{Puschmann} {et~al.}(2010{\natexlab{b}}){Puschmann}, {Ruiz Cobo}, \&
  {Mart{\'{\i}}nez Pillet}}]{pusc10b}
{Puschmann}, K.~G., {Ruiz Cobo}, B., \& {Mart{\'{\i}}nez Pillet}, V.
  2010{\natexlab{b}}, \apjl, 721, L58

\bibitem[{{Rachkowsky}(1967)}]{rach67}
{Rachkowsky}, D.~N. 1967, Izv. Krymsk. Astrofiz. Obs., 37, 56

\bibitem[{{Rempel}(2011)}]{remp11a}
{Rempel}, M. 2011, \apj, 729, 5

\bibitem[{{Rempel}(2012)}]{remp12}
{Rempel}, M. 2012, \apj, 750, 62

\bibitem[{{Rempel} \& {Schlichenmaier}(2011)}]{remp11}
{Rempel}, M. \& {Schlichenmaier}, R. 2011, Living Reviews in Solar Physics, 8,
  3

\bibitem[{{Rempel} {et~al.}(2009{\natexlab{a}}){Rempel}, {Sch{\"u}ssler},
  {Cameron}, \& {Kn{\"o}lker}}]{remp09a}
{Rempel}, M., {Sch{\"u}ssler}, M., {Cameron}, R.~H., \& {Kn{\"o}lker}, M.
  2009{\natexlab{a}}, Science, 325, 171

\bibitem[{{Rempel} {et~al.}(2009{\natexlab{b}}){Rempel}, {Sch{\"u}ssler}, \&
  {Kn{\"o}lker}}]{remp09b}
{Rempel}, M., {Sch{\"u}ssler}, M., \& {Kn{\"o}lker}, M. 2009{\natexlab{b}},
  \apj, 691, 640

\bibitem[{{Riethm{\"u}ller} {et~al.}(2008){Riethm{\"u}ller}, {Solanki}, \&
  {Lagg}}]{riet08}
{Riethm{\"u}ller}, T.~L., {Solanki}, S.~K., \& {Lagg}, A. 2008, \apjl, 678,
  L157

\bibitem[Riethm{\"u}ller et al.(2013)]{riet13} 
{Riethm{\"u}ller}, T.~L., {Solanki}, S.~K., {van Noort}, M., \& {Tiwari}, S.~K.\ 2013, \aap, 
  554, A53 

\bibitem[{{Rimmele}(1995)}]{rimm95}
{Rimmele}, T.~R. 1995, \aap, 298, 260

\bibitem[{{Rimmele} \& {Marino}(2006)}]{rimm06}
{Rimmele}, T. \& {Marino}, J. 2006, \apj, 646, 593

\bibitem[{{Rimmele}(2008)}]{rimm08}
{Rimmele}, T. 2008, \apj, 672, 684


\bibitem[{{Ruiz Cobo} \& {Bellot Rubio}(2008)}]{ruiz08}
{Ruiz Cobo}, B. \& {Bellot Rubio}, L.~R. 2008, \aap, 488, 749

\bibitem[{{Ruiz Cobo} \& {Asensio Ramos}(2013)}]{ruiz13}
{Ruiz Cobo}, B. \& {Asensio Ramos}, A. 2013, \aap, 549, L4

\bibitem[{{Ruiz Cobo} \& {Puschmann}(2012)}]{ruiz12}
{Ruiz Cobo}, B. \& {Puschmann}, K.~G. 2012, \apj, 745, 141

\bibitem[{{Sainz Dalda} \& {Bellot Rubio}(2008)}]{sain08}
{Sainz Dalda}, A. \& {Bellot Rubio}, L.~R. 2008, \aap, 481, L21

\bibitem[{{S{\'a}nchez Almeida} {et~al.}(2007){S{\'a}nchez Almeida},
  {M{\'a}rquez}, {Bonet}, \& {Dom{\'{\i}}nguez Cerde{\~n}a}}]{sanc07}
{S{\'a}nchez Almeida}, J., {M{\'a}rquez}, I., {Bonet}, J.~A., \&
  {Dom{\'{\i}}nguez Cerde{\~n}a}, I. 2007, \apj, 658, 1357

\bibitem[{{Scharmer}(2009)}]{scha09}
{Scharmer}, G.~B. 2009, \ssr, 144, 229

\bibitem[{{Scharmer} {et~al.}(2003){Scharmer}, {Bjelksjo}, {Korhonen},
  {Lindberg}, \& {Petterson}}]{scha03}
{Scharmer}, G.~B., {Bjelksjo}, K., {Korhonen}, T.~K., {Lindberg}, B., \&
  {Petterson}, B. 2003, in Society of Photo-Optical Instrumentation Engineers
  (SPIE) Conference Series, Vol. 4853, Society of Photo-Optical Instrumentation
  Engineers (SPIE) Conference Series, ed. S.~L. {Keil} \& S.~V. {Avakyan},
  341--350

\bibitem[{{Scharmer} {et~al.}(2013){Scharmer}, {de la Cruz Rodriguez},
  {S{\"u}tterlin}, \& {Henriques}}]{scha13}
{Scharmer}, G.~B., {de la Cruz Rodriguez}, J., {S{\"u}tterlin}, P., \&
  {Henriques}, V.~M.~J. 2013, \aap, 553, A63

\bibitem[{{Scharmer} {et~al.}(2002){Scharmer}, {Gudiksen}, {Kiselman},
  {L{\"o}fdahl}, \& {Rouppe van der Voort}}]{scha02}
{Scharmer}, G.~B., {Gudiksen}, B.~V., {Kiselman}, D., {L{\"o}fdahl}, M.~G., \&
  {Rouppe van der Voort}, L.~H.~M. 2002, \nat, 420, 151

\bibitem[{{Scharmer} \& {Henriques}(2012)}]{scha12}
{Scharmer}, G.~B. \& {Henriques}, V.~M.~J. 2012, \aap, 540, A19

\bibitem[{{Scharmer} {et~al.}(2011){Scharmer}, {Henriques}, {Kiselman}, \& {de
  la Cruz Rodr{\'{\i}}guez}}]{scha11}
{Scharmer}, G.~B., {Henriques}, V.~M.~J., {Kiselman}, D., \& {de la Cruz
  Rodr{\'{\i}}guez}, J. 2011, Science, 333, 316

\bibitem[{{Scharmer} {et~al.}(2008){Scharmer}, {Nordlund}, \&
  {Heinemann}}]{scha08}
{Scharmer}, G.~B., {Nordlund}, {\AA}., \& {Heinemann}, T. 2008, \apjl, 677,
  L149

\bibitem[{{Scharmer} \& {Spruit}(2006)}]{scha06}
{Scharmer}, G.~B. \& {Spruit}, H.~C. 2006, \aap, 460, 605

\bibitem[{{Schlichenmaier}(2009)}]{schl09}
{Schlichenmaier}, R. 2009, \ssr, 144, 213

\bibitem[{{Schlichenmaier} {et~al.}(2010){Schlichenmaier}, {Bello
  Gonz{\'a}lez}, {Rezaei}, \& {Waldmann}}]{schl10}
{Schlichenmaier}, R., {Bello Gonz{\'a}lez}, N., {Rezaei}, R., \& {Waldmann},
  T.~A. 2010, Astronomische Nachrichten, 331, 563

\bibitem[{{Schlichenmaier} {et~al.}(1999){Schlichenmaier}, {Bruls}, \&
  {Sch{\"u}ssler}}]{schl99}
{Schlichenmaier}, R., {Bruls}, J.~H.~M.~J., \& {Sch{\"u}ssler}, M. 1999, \aap,
  349, 961

\bibitem[{{Schlichenmaier} {et~al.}(1998{\natexlab{a}}){Schlichenmaier},
  {Jahn}, \& {Schmidt}}]{schl98a}
{Schlichenmaier}, R., {Jahn}, K., \& {Schmidt}, H.~U. 1998{\natexlab{a}},
  \apjl, 493, L121

\bibitem[{{Schlichenmaier} {et~al.}(1998{\natexlab{b}}){Schlichenmaier},
  {Jahn}, \& {Schmidt}}]{schl98b}
{Schlichenmaier}, R., {Jahn}, K., \& {Schmidt}, H.~U. 1998{\natexlab{b}}, \aap,
  337, 897

\bibitem[{{Schlichenmaier} \& {Schmidt}(1999)}]{schl99a}
{Schlichenmaier}, R. \& {Schmidt}, W. 1999, \aap, 349, L37

\bibitem[{{Schlichenmaier} \& {Schmidt}(2000)}]{schl00}
{Schlichenmaier}, R. \& {Schmidt}, W. 2000, \aap, 358, 1122

\bibitem[{{Schlichenmaier} \& {Solanki}(2003)}]{schl03}
{Schlichenmaier}, R. \& {Solanki}, S.~K. 2003, \aap, 411, 257

\bibitem[{{Schmidt} {et~al.}(1992){Schmidt}, {Hofmann}, {Balthasar}, {Tarbell},
  \& {Frank}}]{schm92}
{Schmidt}, W., {Hofmann}, A., {Balthasar}, H., {Tarbell}, T.~D., \& {Frank},
  Z.~A. 1992, \aap, 264, L27

\bibitem[{{Schroeter} {et~al.}(1985){Schroeter}, {Soltau}, \& {Wiehr}}]{schr85}
{Schroeter}, E.~H., {Soltau}, D., \& {Wiehr}, E. 1985, Vistas in Astronomy, 28,
  519

\bibitem[{{Sch{\"u}ssler} \& {V{\"o}gler}(2006)}]{schu06}
{Sch{\"u}ssler}, M. \& {V{\"o}gler}, A. 2006, \apjl, 641, L73

\bibitem[{{Shimizu} {et~al.}(2008){Shimizu}, {Nagata}, {Tsuneta}, {Tarbell},
  {Edwards}, {Shine}, {Hoffmann}, {Thomas}, {Sour}, {Rehse}, {Ito},
  {Kashiwagi}, {Tabata}, {Kodeki}, {Nagase}, {Matsuzaki}, {Kobayashi},
  {Ichimoto}, \& {Suematsu}}]{shim08}
{Shimizu}, T., {Nagata}, S., {Tsuneta}, S., {et~al.} 2008, \solphys, 249, 221

\bibitem[{{Skumanich} {et~al.}(1994){Skumanich}, {Lites}, \& {Mart{\'{\i}}nez
  Pillet}}]{skum94}
{Skumanich}, A., {Lites}, B.~W., \& {Mart{\'{\i}}nez Pillet}, V. 1994, in Solar
  Surface Magnetism, ed. R.~J. {Rutten} \& C.~J. {Schrijver}, 99

\bibitem[{{Skumanich} {et~al.}(1997){Skumanich}, {Lites}, {Martinez Pillet}, \&
  {Seagraves}}]{skum97}
{Skumanich}, A., {Lites}, B.~W., {Martinez Pillet}, V., \& {Seagraves}, P.
  1997, \apjs, 110, 357

\bibitem[{{Solanki}(1987)}]{sola87}
{Solanki}, S.~K. 1987, PhD thesis, PhD thesis No.~8309, ETH, Z{\"u}rich

\bibitem[{{Solanki}(2003)}]{sola03}
{Solanki}, S.~K. 2003, \aapr, 11, 153

\bibitem[{{Solanki} {et~al.}(2010){Solanki}, {Barthol}, {Danilovic}, {Feller},
  {Gandorfer}, {Hirzberger}, {Riethm{\"u}ller}, {Sch{\"u}ssler}, {Bonet},
  {Mart{\'{\i}}nez Pillet}, {del Toro Iniesta}, {Domingo}, {Palacios},
  {Kn{\"o}lker}, {Bello Gonz{\'a}lez}, {Berkefeld}, {Franz}, {Schmidt}, \&
  {Title}}]{sola10}
{Solanki}, S.~K., {Barthol}, P., {Danilovic}, S., {et~al.} 2010, \apjl, 723,
  L127

\bibitem[{{Solanki} \& {Montavon}(1993)}]{sola93}
{Solanki}, S.~K. \& {Montavon}, C.~A.~P. 1993, \aap, 275, 283

\bibitem[{{Solanki} {et~al.}(1994){Solanki}, {Montavon}, \&
  {Livingston}}]{sola94}
{Solanki}, S.~K., {Montavon}, C.~A.~P., \& {Livingston}, W. 1994, \aap, 283,
  221

\bibitem[{{Spruit} \& {Scharmer}(2006)}]{spru06}
{Spruit}, H.~C. \& {Scharmer}, G.~B. 2006, \aap, 447, 343

\bibitem[{{Spruit} {et~al.}(2010){Spruit}, {Scharmer}, \&
  {L{\"o}fdahl}}]{spru10}
{Spruit}, H.~C., {Scharmer}, G.~B., \& {L{\"o}fdahl}, M.~G. 2010, \aap, 521,
  A72

\bibitem[{{Stanchfield} {et~al.}(1997){Stanchfield}, {Thomas}, \&
  {Lites}}]{stan97}
{Stanchfield}, II, D.~C.~H., {Thomas}, J.~H., \& {Lites}, B.~W. 1997, \apj, 477, 485

\bibitem[{{Suematsu} {et~al.}(2008){Suematsu}, {Tsuneta}, {Ichimoto},
  {Shimizu}, {Otsubo}, {Katsukawa}, {Nakagiri}, {Noguchi}, {Tamura}, {Kato},
  {Hara}, {Kubo}, {Mikami}, {Saito}, {Matsushita}, {Kawaguchi}, {Nakaoji},
  {Nagae}, {Shimada}, {Takeyama}, \& {Yamamuro}}]{suem08}
{Suematsu}, Y., {Tsuneta}, S., {Ichimoto}, K., {et~al.} 2008, \solphys, 249,
  197

\bibitem[{{Thomas} \& {Montesinos}(1993)}]{thom93}
{Thomas}, J.~H. \& {Montesinos}, B. 1993, \apj, 407, 398

\bibitem[{{Thomas} \& {Weiss}(2004)}]{thom04}
{Thomas}, J.~H. \& {Weiss}, N.~O. 2004, ARAA, 42, 517

\bibitem[{{Thomas} \& {Weiss}(2008)}]{thom08}
{Thomas}, J.~H. \& {Weiss}, N.~O. 2008, {Sunspots and Starspots} (Cambridge
  University Press)

\bibitem[{{Thomas} {et~al.}(2002){Thomas}, {Weiss}, {Tobias}, \&
  {Brummell}}]{thom02}
{Thomas}, J.~H., {Weiss}, N.~O., {Tobias}, S.~M., \& {Brummell}, N.~H. 2002,
  \nat, 420, 390

\bibitem[{{Title} {et~al.}(1993){Title}, {Frank}, {Shine}, {Tarbell}, {Topka},
  {Scharmer}, \& {Schmidt}}]{titl93}
{Title}, A.~M., {Frank}, Z.~A., {Shine}, R.~A., {et~al.} 1993, \apj, 403, 780

\bibitem[{{Tiwari} {et~al.}(2009){Tiwari}, {Venkatakrishnan}, \&
  {Sankarasubramanian}}]{tiw09b}
{Tiwari}, S.~K., {Venkatakrishnan}, P., \& {Sankarasubramanian}, K. 2009,
  \apjl, 702, L133

\bibitem[{{Tritschler}(2009)}]{trit09}
{Tritschler}, A. 2009, in Astronomical Society of the Pacific Conference
  Series, Vol. 415, The Second Hinode Science Meeting: Beyond Discovery-Toward
  Understanding, ed. B.~{Lites}, M.~{Cheung}, T.~{Magara}, J.~{Mariska}, \&
  K.~{Reeves}, 339

\bibitem[{{Tritschler} {et~al.}(2004){Tritschler}, {Schlichenmaier}, {Bellot
  Rubio}, {KAOS Team}, {Berkefeld}, \& {Schelenz}}]{trit04}
{Tritschler}, A., {Schlichenmaier}, R., {Bellot Rubio}, L.~R., {et~al.} 2004,
  \aap, 415, 717

\bibitem[{{Tsuneta} {et~al.}(2008){Tsuneta}, {Ichimoto}, {Katsukawa}, {Nagata},
  {Otsubo}, {Shimizu}, {Suematsu}, {Nakagiri}, {Noguchi}, {Tarbell}, {Title},
  {Shine}, {Rosenberg}, {Hoffmann}, {Jurcevich}, {Kushner}, {Levay}, {Lites},
  {Elmore}, {Matsushita}, {Kawaguchi}, {Saito}, {Mikami}, {Hill}, \&
  {Owens}}]{tsun08}
{Tsuneta}, S., {Ichimoto}, K., {Katsukawa}, Y., {et~al.} 2008, \solphys, 249,
  167

\bibitem[{{Unno}(1956)}]{unno56}
{Unno}, W. 1956, \pasj, 8, 108

\bibitem[{{van Noort}(2012)}]{van12}
{van Noort}, M. 2012, \aap, 548, A5

\bibitem[{{van Noort} {et~al.}(2013){van Noort}, {Lagg}, {Tiwari}, \&
  {Solanki}}]{downflowpaper}
{van Noort}, M., {Lagg}, A., {Tiwari}, S.~K., \& {Solanki}, S.~K. 2013, \aap,
   submitted

\bibitem[{{Venkatakrishnan} \& {Tiwari}(2010)}]{venk10}
{Venkatakrishnan}, P. \& {Tiwari}, S.~K. 2010, \aap, 516, L5+

\bibitem[{{Westendorp Plaza} {et~al.}(2001){Westendorp Plaza},
  {del Toro Iniesta}, {Ruiz Cobo}, \& {Mart{\'{\i}}nez Pillet}}]{west01a}
{Westendorp Plaza}, C., {del Toro Iniesta}, J.~C., {Ruiz Cobo}, B., \&
  {Mart{\'{\i}}nez Pillet}, V. 2001, \apj, 547, 1148

\bibitem[{{Westendorp Plaza} {et~al.}(1997){Westendorp Plaza}, {del Toro
  Iniesta}, {Ruiz Cobo}, {Martinez Pillet}, {Lites}, \& {Skumanich}}]{west97}
{Westendorp Plaza}, C., {del Toro Iniesta}, J.~C., {Ruiz Cobo}, B., {et~al.}
  1997, \nat, 389, 47


\bibitem[{{Wiehr}(2000)}]{wieh00}
{Wiehr}, E. 2000, \solphys, 197, 227

\bibitem[{{Wiehr} \& {Stellmacher}(1989)}]{wieh89}
{Wiehr}, E. \& {Stellmacher}, G. 1989, \aap, 225, 528

\bibitem[{{Zakharov} {et~al.}(2008){Zakharov}, {Hirzberger}, {Riethm{\"u}ller},
  {Solanki}, \& {Kobel}}]{zakh08}
{Zakharov}, V., {Hirzberger}, J., {Riethm{\"u}ller}, T.~L., {Solanki}, S.~K.,
  \& {Kobel}, P. 2008, \aap, 488, L17

\bibitem[{{Zhang} {et~al.}(2003){Zhang}, {Solanki}, \& {Wang}}]{zhan03}
{Zhang}, J., {Solanki}, S.~K., \& {Wang}, J. 2003, \aap, 399, 755

\end{thebibliography}


\end{document}